\documentclass[preprint,preprintnumbers,amsmath,amssymb]{revtex4} 

\usepackage{graphicx}
\usepackage{dcolumn}
\usepackage{bm}
\usepackage[cp1250]{inputenc}
\usepackage{ifthen} 
\usepackage{amssymb,amstext,amsfonts,amsmath}

\begin{document}

\newcommand{\bk}{\mathbf{k}}
\newcommand{\bq}{\mathbf{Q}}
\newcommand{\bQ}{\mathbf{Q}}
\newcommand{\om}{\overline{m}}
\newcommand{\ua}{\uparrow}
\newcommand{\da}{\downarrow}
\newcommand{\ra}{\rightarrow}
\newcommand{\mb}{\mathbf}
\newcommand{\dg}{^\dagger}
\newcommand{\hw}{\hbar\omega_C}
\newcommand{\eq}{\begin{equation}}
\newcommand{\eqx}{\end{equation}}
\newcommand{\eqn}{\begin{eqnarray}}
\newcommand{\eqnx}{\end{eqnarray}}

\preprint{APS/123-QED}

\title{Statistically-consistent Gutzwiller approach  and its equivalence with the mean-field slave-boson method for correlated systems}

\author{Jakub J\c{e}drak}
\email{j_jedrak@interia.pl}
\affiliation{
Marian Smoluchowski Institute of Physics, Jagiellonian University, \linebreak Reymonta 4, 30-059 Kraków, Poland
}

\author{Jan Kaczmarczyk}
\email{jan.kaczmarczyk@uj.edu.pl}
\affiliation{
Marian Smoluchowski Institute of Physics, Jagiellonian University, \linebreak Reymonta 4, 30-059 Kraków, Poland
}

\author{Jozef Spałek}
 \email{ufspalek@if.uj.edu.pl}
\affiliation{
Marian Smoluchowski Institute of Physics, Jagiellonian University, \linebreak Reymonta 4, 30-059 Kraków, Poland
}
\affiliation{
Faculty of Physics and Applied Computer Science, AGH University of Science and Technology, Reymonta 19, 30-059 Kraków, Poland
}
\date{\today}

\begin{abstract}
We propose a new method of solving a class of mean-field (MF) models, which is based on the Maximum Entropy (MaxEnt) principle with additional constraints included. Next, we show equivalence of our method when applied to the Gutzwiller approximation (GA), with the mean-field slave-boson (SB) formalism (on the example of the single-band Hubbard model). This equivalence provides thus an alternative justification of the results obtained within the SB approach which, however, contains {\it ad hoc} assumptions to position it in agreement with GA. Our approach implies that all predictions of the MF SB method can be obtained in a simpler, transparent, and controllable manner within GA when supplemented with the statistical-consistency conditions. We call the method as the \textit{Statistically-consistent Gutzwiller Approximation} (\textit{SGA}). Explicitly, the present formulation does not require introducing the condensed amplitudes of auxiliary Bose fields, which do not have a direct physical meaning and do not appear in the present formulation.
Although the results of SGA are in the present case equivalent to SB, one can improve them further by utilizing more advanced schemes of calculating averages beyond the standard GA. To illustrate our approach, as well as to outline its advantages over alternative treatments of GA, we select the case of \textit{almost localized Fermi liquid} (ALFL) in two dimensions and analyze it in detail within the tight-binding approximation.
We also comment on significance of our method for describing correlated fermions.
Namely, the reasoning used here can be applied to the corresponding MF treatment of the multiband Hubbard, the periodic Anderson, the $t$-$J$, and the $t$-$J$-$U$ models. In this manner our method may be applied for strongly correlated electron systems, optical lattices, and other related situations.
\end{abstract}


\maketitle

\section{Introduction}
The properties of many-particle systems are frequently determined by a predominance of the interparticle interactions over the single-particle dynamics. Among those are the so-called strongly correlated systems: antiferromagnets \cite{Kastner}, heavy fermions \cite{Gegenwart}, and unconventional superconductors \cite{USC}.
For the case of strongly correlated electron systems the conventional band theory fails, but it is commonly believed, that their satisfactory description from the point of view of quantum statistical physics can be achieved within the parametrized Hubbard \cite{Gutzwiller, Hubbard, Kanamori} or related models ($t$-$J$ \cite{Spalek tJ} or periodic Anderson \cite{Anderson model}). Nonetheless, those models are not as yet solved exactly in the most important cases.
Moreover, for realistic values of the model parameters, the magnitude of electron-electron interaction is comparable to or even much larger than the kinetic energy part. Therefore, also the standard perturbation theory is inapplicable. In such situation, it seems natural to employ variational approach based on trial wave functions. Such approach provides exact upper bounds for the ground-state energy or free energy of the original model.
For the single-band Hubbard model, first such state was proposed by Gutzwiller \cite{Gutzwiller}. The \textit{Gutzwiller wave function} (GWF) had later a number of generalizations, e.g. to multi-band systems \cite{Buenemann Weber, Buenemann Weber Gebhard}, periodic Anderson model \cite{Gebhard}, description of superconductivity in the $t$-$J$ model \cite{ZGRS, Edegger, Gossamer}, and to a time-dependent situation \cite{Time dependent Gutzwiller}.  

Unfortunately, a direct analytic evaluation of the expectation values for GWF is limited to dimensionality $D=1$ \cite{Metzner Vollhardt} and $D = \infty$ \cite{Metzner Vollhardt 2}, (see also \cite{Gebhard older, Hetenyi}), but is not possible as yet for the most important cases of $D=2$ and $3$. Those latter cases (as well as various generalizations of the GWF) may be treated by the Variational Monte Carlo (VMC) techniques as in \cite{Edegger, Becca JPSJ, Capello Becca Fabrizio Sorella Tosatti}, but then only small systems can be studied.
However, for GWF in $D=2$ and $D=3$, approximate analytic expressions for the expectation values may also be obtained \cite{Gutzwiller, Vollhardt}.
Importantly, those treatments of GWF can be, and frequently are, analyzed with the help of an effective single-particle Hamiltonian of the mean-field (MF) character.
This is possible due to the particular form of the Gutzwiller-type variational states, which allows for application of the Wick's theorem. \footnote{Namely, all the proposed Gutzwiller-type variational wave functions are derived from a reference uncorrelated state (eigenstate of a noninteracting Hamiltonian), by acting on the latter with a many-body operator (a correlator) \cite{Gebhard older, Hetenyi}.} This type of approximate treatment of the original variational problem leads to ''physical, but essentially uncontrolled results'' \cite{Gebhard older}. Indeed, on one hand, a quasiparticle picture in the spirit of Landau theory of the Fermi liquid accomplished in this manner is transparent and intuitive. On the other, this approximation is uncontrolled in the sense that it is not guaranteed that the approximate ground-state energy is higher than the exact one. Nonetheless, it should be noted that the hierarchy of ground-state energy values $E_G^{(exact)} \leq E_G^{(GWF)} \leq E_G^{(GA)}$ is observed when comparison of the methods is possible to accomplish, namely for the Hubbard chain $(D=1)$ \cite{Kurzyk}, even when the single-particle wave functions entering $t_{ij}$ and $U$ are optimized. The corresponding $E_G$ values are close as a function of $U/|t|$.

In general situation when implementing MF approach, a direct connection with to the original model is obscured, and the minimization of the approximate ground-state energy can no longer provide the only criterion for the quality of such approximation. Nonetheless, approaches of this kind are still widely used, also in many recent works.
In particular, for the GWF, the approximation named after Gutzwiller (as GA), because of its simplicity and physical clarity, is being used in the context of various versions of the Hubbard model \cite{Buenemann Weber, Buenemann Weber Gebhard, BunemannX}, the related $t$-$J$-$U$ model \cite{Yuan, Heiselberg} or the $t$-$J$ model \cite{ZGRS, Edegger, Marcin 1, Marcin Letters} (in the latter case the effective picture is termed \textit{renormalized mean-field theory}, RMFT \cite{Edegger}). Recently, a combination of Gutzwiller with LDA (Local Density Approximation of Density Functional Theory) has been proposed \cite{Deng}, as well as with RPA (Random Phase Approximation) \cite{Seibold, DiCiolo, Markiewicz}.
Also, closely related to GA are the MF approximations to various versions of the slave-boson (SB) formalism (cf. e.g. \cite{KR}). In all those situations, a description of the system is  based entirely on the effective single-particle Hamiltonian, which is then a starting point of the analysis.

If an approximate treatment of the problems involving variational wave functions purely in terms of an effective single-particle picture is chosen, the task is to solve the effective MF model, i.e. to determine the optimal values of the mean-field parameters appearing in the effective Hamiltonian.
\textit{The principal task of the present paper is to provide an optimal way of performing this procedure.}
In the situation, when we cannot invoke the original variational principle, i.e. minimize the energy difference between approximate and the exact solution, we should search for another criterion of the approximate solution quality.
This aim is achieved by invoking the \textit{maximum entropy} (MaxEnt) principle \cite{Jaynes, Jaynes 2, statistical mechanics}, as  proposed recently in our group \cite{JJJS_arx_0}. The MaxEnt principle is a basis of Bayesian mathematical statistics \cite{Jaynes Prob Logic of Science}. It allows for a construction of the least biased probability distribution on the basis of an incomplete prior information. Therefore, we may hope that the maximum-entropy inference applied to MF models provides the truly optimal solution, at least in that respect.
However, in the case of the MF models the Hamiltonian contains the averages (mean-fields), i.e. implicitly depends on the probability distribution. To deal with such a nonstandard situation, we incorporate additional constraints (with the help of the method of Lagrange multipliers) into the MaxEnt method in such a manner, that the resultant variational formalism automatically preserves the self-consistency of the MF model. Namely, the values of the MF parameters determined from variational procedure are then equal to the average values of the corresponding operators, what is not guaranteed \textit{a priori}. Those additional Lagrange multipliers have then a natural interpretation of the molecular fields. Apart from the formal mathematical motivation for their introduction, they are indispensable in order to achieve a physically-consistent description. We call the resulting method \textit{Statistically-consistent Gutzwiller Approximation} (\textit{SGA}).

In what follows, we work in the nonzero-temperature regime, but the ground-state properties may be always recovered by taking the $T \to 0$ limit. Then our task is to construct, using only the MF effective Hamiltonian, a description of the system in thermal equilibrium and in contact with a particle reservoir. It is important to note that the constraints will be equally valid at $T=0$. Note that GA is devised for $T=0$ (cf. also \cite{Gebhard older, Gebhard}). Therefore, SGA in the present form is valid only for low $T$. However, the GA method extension to $T>0$ can lead to physically important results \cite{JS86}.

To find the optimal probability distribution (corresponding to the equilibrium situation), one has to maximize the entropy augmented with constraints, both those standard, related to the average value of the MF Hamiltonian and particle number operator, as well as the newly introduced in order to preserve the self-consistency of the model. This is equivalent to minimization of the generalized MF grand potential with respect to MF variables, which in the $T \to 0$ limit reduces to finding the minimum of the MF ground-state energy. \footnote{This procedure may be regarded from the point of view of the MaxEnt inference and not the original variational problem. This provides us with a new interpretation of the minimization of the MF thermodynamic potentials, which is also consistent with the interpretation of the MF models in terms of the Landau theory of phase transitions.}

The method we use is simple, transparent, and of general applicability, in the sense, that it may be applied to \textit{any} MF model, also to any effective Hamiltonian resulting from different approximate treatments of GWF. In particular, when applied to the simplest GA, this method yields results equivalent (see Sec. \ref{Equi}) to the saddle-point MF approximation of the SB formalism of Kotliar and Ruckenstein \cite{KR}. Therefore, our paper adds to the discussion of SB-GA equivalence addressed in the series of papers \cite{Gebhard older, Gebhard, Buenemann Gebhard}. However, the equivalence is present not only in the simplest situation, but also for the GA and MF SB treatment of multiband Hubbard model (see Sec. \ref{sec:appGen}) and for the corresponding spin-rotational-invariant versions of the SB/GA formalism (see Secs. \ref{sec:appGen} and \ref{sec:extendC}). Here, for simplicity, we concentrate mainly on the case of GA for the single-band Hubbard model, as given e.g. in \cite{Vollhardt}.

Our approach provides an alternative justification and derivation of the MF SB formalism. Namely, all features of the latter may be obtained in an alternative, simpler manner and without introducing the condensed-Bose fields. SGA may be also applied to construct and solve effective single-particle models corresponding to more complicated versions of approximate treatments of GWF, e.g. those including $1/D$ corrections \cite{Gebhard older} or inter-site correlations \cite{Fukushima}, which cannot be reproduced by any form of the SB formalism. By using generalized schemes of calculating expectation values in GWF \cite{Fukushima} and the optimal way of solving the resulting MF model, we can systematically improve the GA solution. Doing so, one may hope to obtain solutions of similar quality to those offered by GWF (via VMC calculations), but within a procedure not limited to small systems.

Finally, let us note that the violation of the upper bound for exact free- or ground-state energies of the original (e.g. Hubbard) model is connected with the nature of GA itself, and not with the method we propose.

It should be noted here, that several examples of self-consistent, zero-temperature variational MF approaches, some of which  are equivalent to our approach if the $T\to 0$ limit is taken, are existing already in the literature \cite{Gebhard, Buenemann Gebhard Thul, Ogata Himeda, LiZhouWang, Wang F C Zhang, Fazekas}. Yet, none of them is based on the MaxEnt principle and none of them tackles equivalence with the SB formalism. Also, the MaxEnt-based MF approach was developed in a different form in Refs. \cite{Argentynczycy}. Furthermore, frequently the MF models are solved by using basic self-consistency conditions (in the context of superconductivity termed as the Bogoliubov-de Gennes equations), and no variational procedure is invoked \cite{Sigrist, Didier, Marcin 1, Marcin Letters}. Finally, other methods are also used, being variational in nature, but not equivalent to the MaxEnt treatment \cite{Vollhardt, Buenemann Weber, Buenemann Weber Gebhard}. Therefore, it seems there is no consensus between different groups working in the field on how to solve the MF models resulting from GA. In brief, the existing number of papers, as well as our earlier works \cite{JJJS_arx_0, Jedrak Spalek} motivated us to systematize the approach, at least from the statistical-physics point of view.

The structure of the paper is as follows. In Sec. \ref{sec:formalism} we review briefly the GA formalism. In Sec. \ref{GA2} we discuss the MaxEnt approach, in which GA is supplemented with the self-consistency constraints. This is the way we implement the MaxEnt principle. In Sec.~\ref{sec:interim} we discuss the differences between GA and SGA. In Sec. \ref{SB} we review briefly the SB formalism and carry out a detailed analysis of the SGA and SB formalisms equivalence. In Sec. \ref{sec:appGen} we comment on a generalization of the SGA approach to spin-rotationally invariant and multiorbital situations. In Sec. \ref{sec:numerics} we illustrate the SGA method by considering the single-band Hubbard model. We analyze nontrivial magnetic features and compare our results with the GA approach to emphasize the novel features of the SGA method. In Sec. \ref{sec:extend} we outline some possibilities of extending our approach and comment on its significance for the description of other strongly-correlated  systems. In Sec. \ref{sec:summary} we summarize our results. Finally, in Appendix \ref{appHF} we discuss the Hartree-Fock limit of our method on the example of the Stoner theory.

\section{Model and brief summary of Gutzwiller ansatz (GA)} \label{sec:formalism}

We start from the single-band Hubbard Hamiltonian, which has the form
\begin{equation}
\hat{H} = \sum_{ij\sigma}{} t_{ij} c^{\dag}_{i\sigma}c_{j\sigma} + U\sum_{i} \hat{n}_{i \uparrow} \hat{n}_{i \downarrow},
\label{Hubbard Hamiltonian}
\end{equation}
where the first term expresses particle hopping between the sites $i$ and $j$ (with the hopping amplitude $t_{ij}$) and the second describes the intra-atomic repulsive interaction characterized by the Hubbard parameter $U$.
In the following $\Lambda$, $N_{\uparrow}$, ($N_{\downarrow}$), and $D$ denote the number of lattice sites, of spin up (down) electrons, and of double occupied sites, respectively. Also, $n_{\sigma} \equiv N_{\sigma}/\Lambda$ for $\sigma = \uparrow, \downarrow$ and $D/\Lambda \equiv d^2$ (this quantity is identical with the quantity $d$ of Ref.~\onlinecite{Vollhardt} and $d^2$ of Ref. \onlinecite{KR}).

We summarize here the standard \footnote{The word ''standard'' refers to e.g. expressions for the approximate ground-state energy. The method of solution presented here is the simplest one, but it cannot be called standard, as in the literature there are many different ways of solving the MF model resulting from GA.} GA \cite{Gutzwiller} following the notation of Ref. \onlinecite{Vollhardt}. The Gutzwiller trial state $| \psi \rangle $ is derived from an uncorrelated, normalized single-particle state $|\psi_0 \rangle $ by suppressing the weight of those components of the latter, which correspond to one or more doubly occupied sites. In the simplest case, $ | \psi \rangle $ depends on a single variational parameter $g$, i.e. the many-body trial wave function is postulated of the form
\begin{equation}
| \psi \rangle = \prod_{i} [1 - (1 - g)\hat{n}_{i \uparrow} \hat{n}_{i \downarrow} ] | \psi_0 \rangle \equiv \hat{P}_G |\psi_0 \rangle.
\label{Psi Gutzwiller}
\end{equation}
In the present paper $| \psi_0 \rangle$ represents an ordinary Fermi sea, although it may be magnetically polarized (also more complicated uncorrelated states exhibiting e.g. antiferromagnetic and/or superconducting order, can be considered \cite{Edegger}).
Using the projection (\ref{Psi Gutzwiller}), one may try to evaluate the expectation value of the Hamiltonian (\ref{Hubbard Hamiltonian}), i.e. $ \langle \psi|\hat{H}| \psi\rangle/\langle \psi | \psi\rangle $. However, this is a nontrivial task and to deal with it we have to introduce further approximations, which will not be discussed here. In result, following Gutzwiller \cite{Gutzwiller}, we obtain a relatively simple formula for the ground-state energy, which for the translationally invariant state reads
\begin{equation}
 \frac{\langle \psi|\hat{H}| \psi\rangle}{\langle \psi | \psi\rangle} \approx E_{g}/\Lambda = q_{\uparrow}(d, n_{\uparrow}, n_{\downarrow}) \overline{\epsilon}_{\uparrow} + q_{\downarrow}(d, n_{\uparrow}, n_{\downarrow}) \overline{\epsilon}_{\downarrow} + U d^2.
\label{E_g Gutzwiller}
\end{equation}
In the above, the quantity
\begin{equation}
 q_{\sigma}(d, n_\ua, n_\da) = \frac{\{ [ (n_{\sigma} - d^2)(1 - n_{\sigma} - n_{\bar{\sigma}} + d^2 ) ]^{1/2} + d [(n_{\bar{\sigma}} - d^2)]^{1/2}\}^2 }{n_{\sigma}(1 - n_{\sigma})},
\label{q_sigma Gutzwiller}
\end{equation}
has an interpretation of the band narrowing (renormalization) factor and
\begin{equation}
\overline{\epsilon}_{\sigma} = \Lambda^{-1} \langle \psi_{0} | \sum_{ij} t_{ij} c^{\dag}_{i\sigma}c_{j\sigma}| \psi_{0} \rangle = \Lambda^{-1} \sum_{\mathbf{k}} \epsilon_\mathbf{k},
\label{epsilon_sigma Gutzwiller}
\end{equation}
where the $\bk$-summation is taken over the filled part of the bare band with spin $\sigma$ and $\overline{\epsilon}_{\sigma}$ is an average bare band energy per site for particles of spin $\sigma = \pm 1$. It is also convenient to change variables from $n_{\sigma}$ to $n \equiv \sum_{\sigma}n_{\sigma}$ and $m \equiv \sum_{\sigma}\sigma n_{\sigma}$ representing the band filling and magnetic moment (spin-polarization) per site, respectively. It is important to note, that due to the approximate evaluation of the l.h.s. of Eq. (\ref{E_g Gutzwiller}) it is not guaranteed that $E_{g}$ is higher then the exact ground-state energy of the Hubbard model. Also, Eq. (\ref{E_g Gutzwiller}) may be interpreted as an expectation value of an effective single-particle Hamiltonian, $\hat{H}_{GA}$, evaluated with respect to $| \psi_{0} \rangle$, e.g.
\begin{equation}
 E_{g} = \langle \psi_0|\hat{H}_{GA}| \psi_0 \rangle.
\label{various expectation values of}
\end{equation}
From Eqs. (\ref{E_g Gutzwiller})-(\ref{epsilon_sigma Gutzwiller}) it follows directly that
\begin{eqnarray}
\hat{H}_{GA}(d, n, m) &= & \sum_{ij\sigma} q_{\sigma}(d, n, m) t_{ij} c^{\dag}_{i\sigma}c_{j\sigma} - \sum_{i\sigma } \sigma h c^{\dag}_{i\sigma}c_{i\sigma} + \Lambda U d^2 \nonumber \\
& = & \sum_{\mathbf{k}\sigma} \big(q_{\sigma}(d, n, m) \epsilon_{\mathbf{k}} - \sigma h\big) c^{\dag}_{\mathbf{k}\sigma}c_{\mathbf{k}\sigma} + \Lambda U d^2,
\label{MF Gutzwiller Hamiltonian}
\end{eqnarray}
where the Zeeman term was introduced explicitly (with the reduced magnetic field $h \equiv g \mu_B H_a$). Furthermore, $| \psi_0 \rangle $ is chosen to be the ground state of $\hat{H}_{GA}$, $\hat{H}_{GA} | \psi_0 \rangle = E_g | \psi_0 \rangle$. Thus, the Gutzwiller approximation can be alternatively introduced as based on the effective quasiparticle Hamiltonian (\ref{MF Gutzwiller Hamiltonian}) \cite{JS86}. This approach is termed \textit{renormalized mean-field theory} (RMFT) \cite{ZGRS, Edegger}, as the Hamiltonian (\ref{MF Gutzwiller Hamiltonian}) contains renormalized (by $q_\sigma$) \textit{bare} hopping integral $t_{ij}$ (or single-particle energy $\epsilon_\bk$).

Hamiltonian $\hat{H}_{GA}$ depends in a non-Hartree-Fock manner on the parameters $n$, $m$, and $d$, the values of which are not determined as yet. The first two of them have the meaning of expectation values of single particle operators, i.e. $n = N/\Lambda$ and $ m = M/\Lambda$, where
\begin{equation}
N \equiv \langle \hat{N} \rangle = \sum_{\mathbf{k} \sigma} \langle c^{\dag}_{\mathbf{k} \sigma}c_{\mathbf{k} \sigma} \rangle, ~~~~~ M \equiv \langle \hat{M} \rangle = \sum_{\mathbf{k} \sigma} \sigma \langle c^{\dag}_{\mathbf{k} \sigma} c_{\mathbf{k} \sigma} \rangle.
\label{} \label{eq:sc1}
\end{equation}

Although the Gutzwiller approach was devised for zero temperature, we may still construct the partition function and the (generalized) grand-potential functional $\mathcal{F}^{(GA)}$
\begin{equation}
 \mathcal{F}^{(GA)}   =  -\frac{1}{\beta } \sum_{ \mathbf{k} \sigma }  \ln [1 + e^{-\beta E^{(GA)}_{\mathbf{k} \sigma} }]   +  \Lambda U d^2,
\label{SB SP MF GP simpler Gutzwiller bis}
\end{equation}
with the quasiparticle energies
\begin{equation}
E^{(GA)}_{\mathbf{k} \sigma} = q_{\sigma} \epsilon_{\mathbf{k}}  -  \sigma h   - \mu.
\label{E_k standard Gutzwiller}
\end{equation}
These steps are taken to compare results of a particular way of solving the Gutzwiller approach (note the superscript GA in the above equations) with the generalized Gutzwiller+MaxEnt approach introduced next. Explicitly, within GA solution one minimizes the ''Landau functional'' (\ref{SB SP MF GP simpler Gutzwiller bis}) with respect to the variational parameter $d$, which leads to the condition
\begin{equation}
\frac{ \partial \mathcal{F}^{(GA)}}{\partial d} = 2 \Lambda U d   +  \sum_{ \mathbf{k} \sigma } \frac{\partial  q_{\sigma}}{\partial d} f(E^{(GA)}_{\mathbf{k} \sigma})  \epsilon_{\mathbf{k}} = 0,
\label{der mathF ksi explicite standard}
\end{equation}
with $f(E)$ being the Fermi-Dirac distribution. This equation is supplemented with the self-consistent equations. First, magnetization $m$ is not treated as a variational parameter and consequently, its value is determined from the defining (self-consistent) equation (\ref{eq:sc1}), namely
\begin{equation}
m   = \frac{1}{\Lambda}\sum_{ \mathbf{k} \sigma } \sigma  f(E^{(GA)}_{\mathbf{k} \sigma}).
\label{der mathF lambda m explicite bis}
\end{equation}
Second, the chemical potential is determined from the particle-number conservation, i.e.
\begin{equation}
n   = \frac{1}{\Lambda}\sum_{ \mathbf{k} \sigma } f(E^{(GA)}_{\mathbf{k} \sigma}).
\label{der mathF lambda m explicite2}
\end{equation}
Thus we see, the GA solution contains a mixture of self-consistent equations for $m$ and $\mu$ and a variational minimization of $d$. Eqs. (\ref{der mathF ksi explicite standard})-(\ref{der mathF lambda m explicite2}) form a complete set for $d$, $m$, and $\mu$, which is solved numerically. The above equations express the way of solving the Gutzwiller approximation (GA) used for comparison with SGA; it is used frequently e.g. in the context of the $t$-$J$ model \cite{Didier, Marcin 1, Marcin Letters}. This formulation differs from that of Ref. \onlinecite{Vollhardt}.

Note, that the nonzero temperature formalism presented here, in the $\beta \to \infty$ limit, is fully equivalent to the original Gutzwiller approach devised for $T=0$.

\section{Gutzwiller approximation combined with Max-Ent approach: SGA} \label{GA2}

\subsection{Motivation for the approach}

\begin{figure}
\begin{center}
\includegraphics[height=8cm]{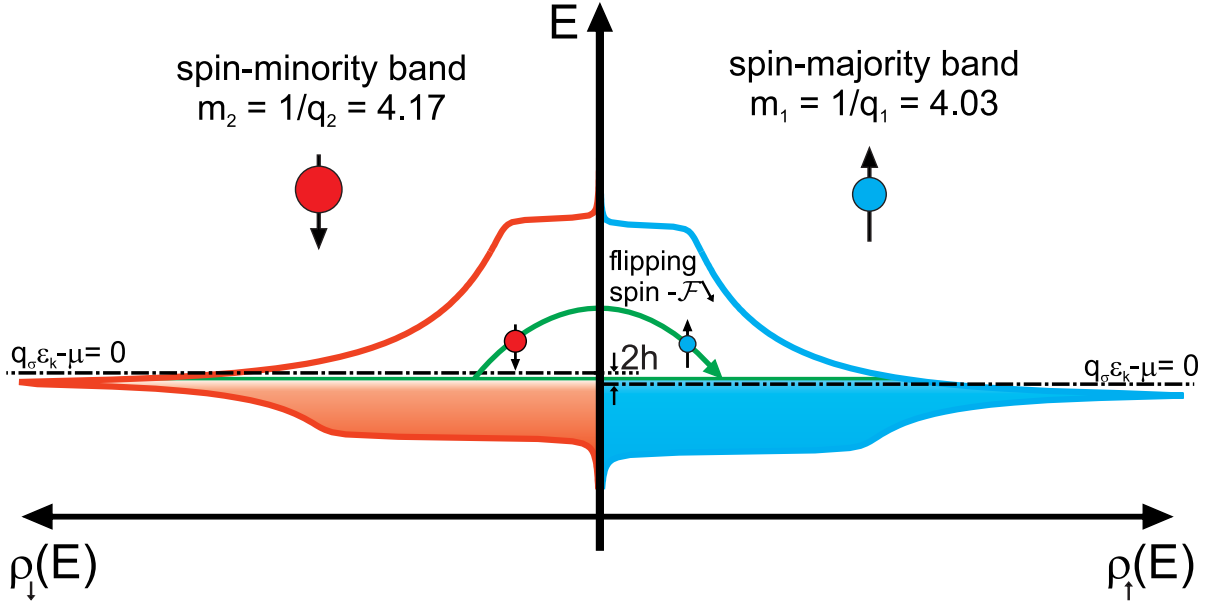}
\end{center}
\caption{(Color online) Spin-resolved density of states for the spin-majority ($\sigma = \ua$) and the spin-minority ($\sigma = \da$) subbands calculated in the standard Gutzwiller approximation. The dot-dashed lines show the reference energy (defined by $q_\sigma \epsilon_\bk - \mu = 0$). Those points of the subbands are shifted by the Zeeman spin splitting $2h$. The calculations were performed for $U=12$ and $h=0.05$ for a two-dimensional band. The quantities $m_1$ and $m_2$ are the corresponding spin-dependent mass-enhancement factors. For details of the numerical procedure see Sec. \ref{sec:numerics}.}
\label{figDOS1}
\end{figure}

To demonstrate directly that the basic method of solving GA summarized in Sec. \ref{sec:formalism} represents not fully-optimized approach, we analyze the ground state in the general case, i.e. when $q_\sigma$ depends on the spin polarization $m$. Then, it is straightforward to show that the derivative $ \partial \mathcal{F} / \partial m = \sum_{\bk \sigma} \frac{\partial q_\sigma(d, n, m)}{\partial m} \epsilon_\bk f(E_{\bk \sigma}) \neq 0$, which physically means that by transferring a small number of particles from one spin-subband to the other (i.e. by changing spin polarization $m$), we observe a decrease in the total energy of the system (see Fig. \ref{figDOS1} for illustration). We may understand this decrease intuitively by noting that the spin transfer process between the subbands leads not only to a change in the energy level occupation (as would be in the standard case), but also to an alteration of the renormalization factor $q_\sigma(d, n, m)$ for  \textit{all} the single particle energy levels. Such instability is present only if the Gutzwiller factors depend explicitly on the spin polarization $m$.

Within our method, we treat $m$ and, other mean-fields as variational parameters, with respect to which the appropriate grand-potential (Landau) functional is minimized. To carry out the procedure, we introduce constraints as discussed next.

\subsection{Formal structure of SGA} \label{sec:formalstructure}

On the technical level, a direct minimization of $\mathcal{F}^{(GA)}$ with respect to $m$ would lead to violation of the self-consistency equation (\ref{der mathF lambda m explicite bis}). Therefore, in order to preserve the self-consistency, additional constraint on $m$ has to be imposed by means of the Lagrange-multiplier method. Analogously, we introduce the constraint on $n$. In general, there should be a constraint for each mean field appearing explicitly in a non-HF manner in the effective MF Hamiltonian (also, for e.g. not included here the staggered magnetization and the pairing amplitude). Here, $m$ and $n$ appear in $\hat{H}_{GA}$ via $q_\sigma(d, n, m)$. The presence of those constraints leads to redefinition of the Hamiltonian (\ref{MF Gutzwiller Hamiltonian}), according to the prescription
\begin{equation}
 \hat{H}_{\lambda} \equiv \hat{H}_{GA} - \lambda_{m}(\hat{M} - M) - \lambda_{n}(\hat{N} - N).
\label{sdm Hamiltonian}
\end{equation}
The Lagrange multipliers $\lambda_{m}$ and $\lambda_{n}$ play the role of (homogenous) molecular fields, which are coupled to the spin polarization and the total charge, respectively (the general, inhomogeneous case can be treated analogously). Similar terms are present in some papers \cite{Gebhard, Buenemann Gebhard Thul, LiZhouWang} and absent in others (for the latter cf. treatment in Ref. \onlinecite{Vollhardt} and in those on application of RMFT to $t$-$J$ model \cite{Edegger, Didier, Marcin 1, Marcin Letters}).
On the contrary, the variational parameter $d$ is not an average value of any operator appearing in $ \hat{H}_{GA} - \mu \hat{N}$, and as such, does not require any self-consistency-preserving constraint.

\begin{figure}
\begin{center}
\includegraphics[height=16cm]{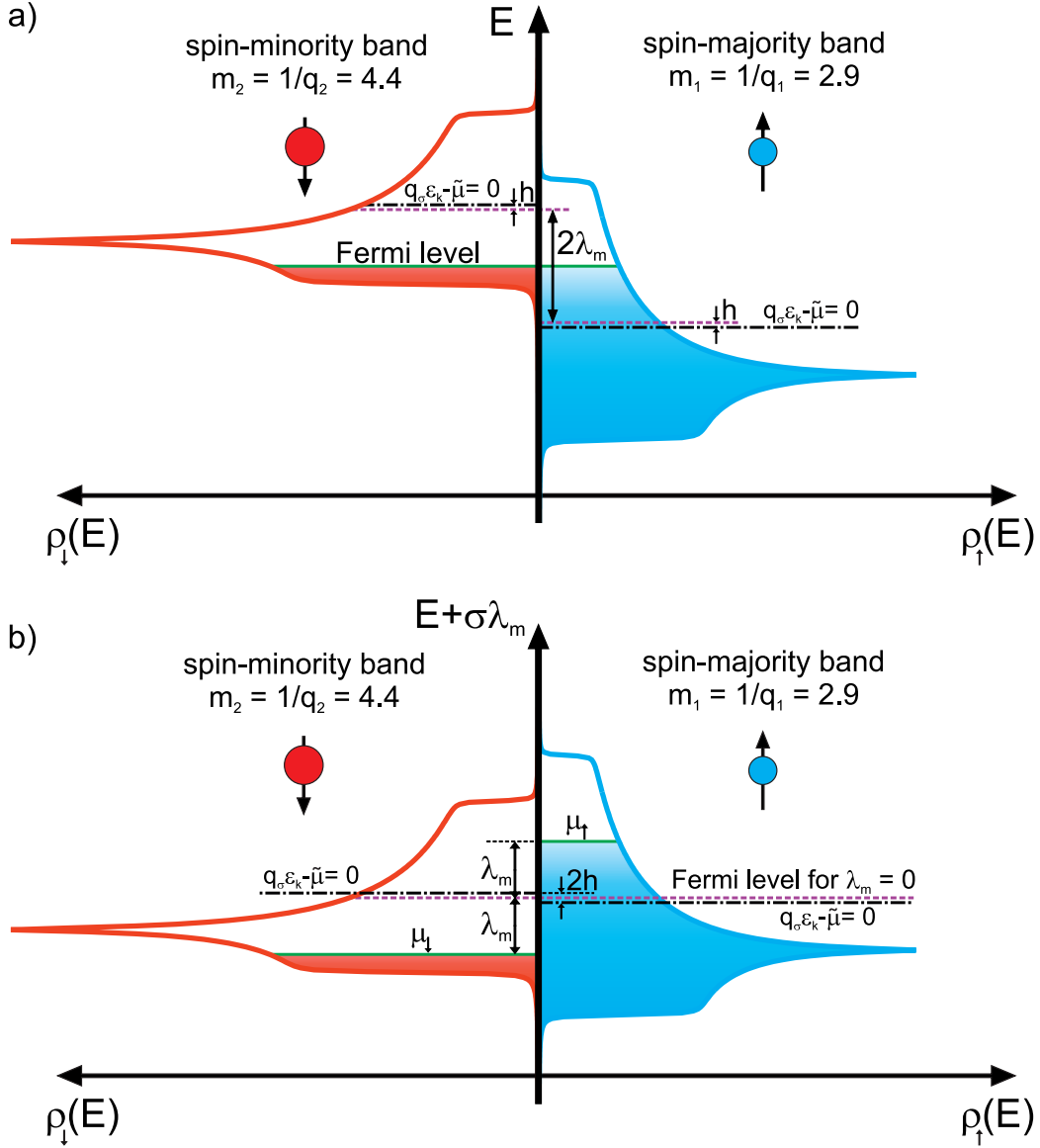}
\end{center}
\caption{(Color online) Density of states for the spin-majority ($\sigma = \ua$) and spin-minority ($\sigma = \da$) subbands obtained within SGA. The Fermi liquid can be viewed equivalently as either (a) with a single chemical potential or (b) with different effective chemical potentials $\mu_\sigma \equiv \tilde{\mu} + \sigma \lambda_m$. The dashed line in (b) shows the Fermi level if we put $\lambda_m = 0$. The dot-dashed lines show the reference energy (defined by $q_\sigma \epsilon_\bk - \tilde{\mu} = 0$). Those points are now shifted not only by $2h$ as previously, but by $2h+2\lambda_m$, which leads to the much greater mass-splitting than in the standard Gutzwiller case (cf. Fig. \ref{figDOS1}). The calculations were performed for the parameter values as in Fig.~\ref{figDOS1}. The numerical procedure is detailed in Sec. \ref{sec:numerics}.}
\label{figDOS2}
\end{figure}

Next, we construct the generalized grand-potential functional $\mathcal{F}$ for the effective Hamiltonian (\ref{sdm Hamiltonian}),
\begin{equation}
\mathcal{F}^{(SGA)} \equiv -\beta^{-1}\ln \mathcal{Z}_{\lambda}, ~~~ ~~~~ \mathcal{Z}_{\lambda} \equiv \text{Tr}[\exp \big(-\beta ( \hat{H}_{\lambda} - \mu \hat{N})\big)].
\label{mathcal F functional}
\end{equation}
Explicitly, we have
\begin{equation}
\mathcal{F}^{(SGA)} = -\frac{1}{\beta } \sum_{ \mathbf{k} \sigma } \ln [1 + e^{-\beta E^{(SGA)}_{\mathbf{k} \sigma} }] + \Lambda( \lambda _{n} n + \lambda _{m} m + U d^2).
\label{SB SP MF GP simpler Gutzwiller0}
\end{equation}
Note that the definition of $\mathcal{F}^{(SGA)}$ is based on $\hat{H}_{\lambda}$, not on $\hat{H}_{GA}$. The quasiparticle energies are thus defined in the form
\begin{equation}
E^{(SGA)}_{\mathbf{k} \sigma} = q_{\sigma} \epsilon_{\mathbf{k}}  - \sigma (h + \lambda_{m}) - \tilde{\mu},
\label{E_k Gutzwiller}
\end{equation}
with $\tilde{\mu} \equiv \mu + \lambda_n$ as shifted chemical potential and $h+\lambda_m$ as an effective magnetic field. All the averages appearing above are defined with the help of the following density operator
\begin{equation}
\hat{\rho}_{\lambda} = \mathcal{Z}_{\lambda}^{-1} \exp \big(-\beta ( \hat{H}_{\lambda} - \mu \hat{N})\big)
\label{correct MF density matrix}
\end{equation}
in a standard manner, i.e. $A \equiv \langle \hat{A} \rangle = \text{Tr}[ \hat{A}\hat{\rho}_{\lambda}] $. The equilibrium values of the mean fields and the Lagrange multipliers are regarded as those, which are optimal from the point of view of the MaxEnt approach. They are obtained from the necessary conditions for $\mathcal{F}$ to have a minimum subject to constraints, i.e.
\begin{equation}
\frac{\partial \mathcal{F}}{\partial \vec{A}} = 0, ~~~~~ \frac{\partial \mathcal{F}}{\partial \vec{\lambda}} =0, ~~~~~ \frac{\partial \mathcal{F}}{\partial d} = 0.
\label{derivative of mathcalF A, lambda}
\end{equation}
In the above equations: $ \partial \mathcal{F} / \partial \vec{A} \equiv \nabla_{A} \mathcal{F} $, etc., and by $\vec{A}$, $\vec{\lambda}$ we denote respectively the sets of the mean fields and of Lagrange multipliers; explicitly: $\vec{A} = (n, m)$ and $\vec{\lambda} = (\lambda_n, \lambda_m)$. Needless to say, that the conditions $\partial \mathcal{F} / \partial \vec{\lambda} = 0$ guarantee the realization of the self-consistent equations automatically.

In effect, the following variables are to be determined from the variational minimization procedure: $d$, $m$, $\lambda_{m}$, $\lambda_{n}$, and $\mu$, with $n$ being fixed. The presence of both $\lambda_{n}$ and $\mu$ at the same time is necessary: the former ensures a self-consistent way of evaluating $n$, whereas the latter fixes $n$ at a desired value. The physical meaning of $\lambda_m$ is illustrated in Fig. \ref{figDOS2}. Namely, $\lambda_m$ optimizes the free energy by allowing for Fermi-level mismatch between the spin-subbands to readjust. The choice of $(n , m)$ instead of $(n_{\uparrow}, n_{\downarrow})$ is more convenient within the grand-canonical formalism. However, one may go back to the original mean fields $(n_{\uparrow}, n_{\downarrow})$. Then, the molecular fields transform accordingly, i.e.
\begin{equation}
(\lambda_{n}, \lambda_{m}) \to (\lambda_{\uparrow}, \lambda_{\downarrow}) \equiv (\lambda_{n} + \lambda_{m}, \lambda_{n} - \lambda_{m}).
\label{transformation of molecular fields}
\end{equation}
The MF thermodynamics is constructed by defining the grand potential $\Omega(T, V, \mu)$ from the generalized grand-potential functional $\mathcal{F}$, evaluated for the optimal values of all parameters (i.e. the solutions of Eqs. (\ref{derivative of mathcalF A, lambda})), and has the form
\begin{equation}
\Omega(T, V, \mu, h) = \mathcal{F}(T, V, \mu, h; \vec{A}_0(T, V, \mu, h), \vec{\lambda}_0(T, V, \mu, h), d_0(T, V, \mu, h)).
\label{Thermo potentials}
\end{equation}
In the above formula $\vec{A}_0(T, V, \mu, h)$, $\vec{\lambda}_0(T, V, \mu, h)$, and $d_0(T, V, \mu, h)$ denote the equilibrium values of the mean-fields, the Lagrange multipliers, and the double occupancy respectively. Consequently, the free energy is defined as $F = \Omega + \mu N$. Note that $\mu$, not $\tilde{\mu} \equiv \mu + \lambda_n$ is present in the above formulas. The equilibrium thermodynamic potentials do not depend on mean-fields or molecular fields, as the latter are removed in the process of the corresponding functional minimization. Therefore, the quantity $\mu$ plays the role of the thermodynamic chemical potential entering in the relations
\eq
\frac{\partial \Omega }{ \partial \mu}  =  -N, \qquad \frac{\partial F }{ \partial N} = \mu. \label{eq:therrel}
\eqx
Parenthetically, if we disregarded $\lambda_n$ (putting $\lambda_n=0$), then the condition $\partial \mathcal{F}/\partial n = 0$ should not be used. In such scheme, the values of the quantities $m$, $\lambda_m$, $d^2$ would be the same, but the relations (\ref{eq:therrel}) would not be fulfilled, and for fixing $n$ the s-c condition should be utilized.

\section{An interim summary: GA vs. SGA} \label{sec:interim}

As follows from earlier discussion, if GA is implemented without the molecular fields ($\vec{\lambda}$), magnetization $m$ cannot be treated as a variational parameter. In order to obtain the correct MF thermodynamics, the functional (\ref{SB SP MF GP simpler Gutzwiller0}) must be minimized also with respect to $m$. Otherwise, inconsistencies in the statistical description of such essentially non-Hartree type of MF model appear even at the level of the first derivatives of the respective thermodynamic potential \cite{JJJS_arx_0} (e.g. $\partial \Omega / \partial h \neq -M$ or $\partial F / \partial h \neq - M$). In other words, the inclusion of the constraints in (\ref{sdm Hamiltonian}) represents not only a possible approach, but first of all it is required in reaching statistical consistency of the results.
A clear example of its indispensability is also the study of the Fulde-Ferrell-Larkin-Ovchinnikov (FFLO) superconducting phase \cite{JKJS} within a Gutzwiller scheme performed by us recently, in which if we disregarded MaxEnt constraints, it would lead to nonphysical results such as a jump in the free energy at the BCS-FFLO phase transition.
In Sec. \ref{sec:numerics}, on example of the GA for Hamiltonian (\ref{Hubbard Hamiltonian}), we show explicitly that these two methods of approach, i.e. either by following the present SGA procedure and treating $m$ as a variational parameter (\textit{var}) or by regarding it as a parameter determined self-consistently as in the basic GA approach (\textit{s-c}), yield quite different results.

A methodological remark is in order at this point. Namely, it may happen that for some MF model the results of the non-variational self-consistent treatment are in better agreement with the experimental results than those of the MaxEnt-based variational method. Still, it does not justify the former way of approach and the agreement should be regarded as accidental, since it is inconsistent with the fundamental principles of mathematical statistics \cite{Jaynes, Jaynes 2}.

Within the simplest MF scheme, e.g. the Hartree-Fock approximation (cf. Appendix \ref{appHF}), the single-particle energies are modified in an additive manner, i.e. according to $\varepsilon \to \varepsilon + b$, where $b$ is some term depending on mean-fields. Only in such case the unwary variational approach (i.e. that without constraints) yields proper self-consistent results (then explicitly $\lambda_n = \lambda_m = 0$, cf. Appendix \ref{appHF} for details). On the other hand, the original Gutzwiller approximation may be viewed as a 'multiplicative' renormalization of bare single-particle energies according to $\varepsilon \to q\varepsilon $. However, in such case, without introduction of the self-consistency conditions in (\ref{sdm Hamiltonian}), we may not regard mean-fields as variational parameters, as they must be determined from non-variational procedure by making use of self-consistency equations only. Then, the corresponding grand potential or energy functionals are allowed to be minimized only with respect to the variables, which, like $d$, are not expectation values of any single-particle operator present in the MF Hamiltonian. To obtain a complete and consistent statistical description of any MF model, we have to introduce additional constraint terms, so the 'multiplicative' renormalization  is supplemented also with the 'additive' one, i.e. $\varepsilon \to q\varepsilon + \lambda$.

In the next Sec. we summarize briefly the slave boson formalism (SB) and then proceed with its equivalence with the method developed in Sec. \ref{GA2}.

\section{Slave boson formalism and the equivalence with SGA \label{SB}}

\subsection{Saddle - point approximation in the slave boson approach}

Below we briefly recall the main features of the formalism of Ref. \onlinecite{KR}. Our starting point is once again the Hamiltonian (\ref{Hubbard Hamiltonian}).
The main idea of SB method is that of an enlargement of the Fock space by replacing the original fermionic space of fermion operators $c^{\dag}_{i\sigma}$ ($c_{i\sigma}$) with that of auxiliary fermion $f^{\dag}_{i\sigma}$ ($f_{i\sigma}$) and boson operators. Explicitly, $e^{\dag}_{i}$ ($e_{i}$), $p^{\dag}_{i\sigma}$ ($p_{i\sigma}$) and $d^{\dag}_{i}$ ($d_{i}$) are extra boson fields which create (annihilate) the empty, singly and doubly occupied states out of the postulated vacuum state, respectively. Such an enlarged local Hilbert space contains in an obvious manner nonphysical states. To eliminate them the following local constraints (for each site $i$) are introduced, namely
\begin{equation}
\hat{Q}^{(1)}_{i} =  \sum_{\sigma} p^{\dag}_{i\sigma} p_{i\sigma} + e^{\dag}_{i} e_{i} + d^{\dag}_{i} d_{i} - 1 = 0,
\label{SB completeness}
\end{equation}
\begin{equation}
\hat{Q}^{(2)}_{i\sigma} = f^{\dag}_{i\sigma} f_{i\sigma} - p^{\dag}_{i\sigma} p_{i\sigma} - d^{\dag}_{i} d_{i} = 0.
\label{SB equivalence}
\end{equation}
Eq. (\ref{SB completeness}) reflects the completeness relation, whereas (\ref{SB equivalence}) equates the two ways of counting electrons via fermionic or bosonic representations, respectively. Hamiltonian (\ref{Hubbard Hamiltonian}) written in terms of the new operators, reads
\begin{equation}
\hat{H} = \sum_{ij\sigma} t_{ij} \hat{z}^{\dag}_{i\sigma}\hat{z}_{j\sigma} f^{\dag}_{i\sigma}f_{j\sigma} + U\sum_{i} d^{\dag}_{i} d_{i},
\label{SB Hubbard Hamiltonian}
\end{equation}
where $\hat{z}_{i\sigma} = e^{\dag}_{i} p_{i\sigma} + p^{\dag}_{i\bar{\sigma}} d_{i}$. We see that the Hubbard term $U\sum_{i} \hat{n}_{i \uparrow} \hat{n}_{i \downarrow}$ is expressed in terms of bosonic operators and has a simple (single-particle) form. In contrast, the kinetic-energy part is a quite complicated expression and contains interaction between the auxiliary fermions and bosons.

As a next step, the partition function is written as the functional integral over the Bose and Fermi fields. The constraints (\ref{SB completeness}), (\ref{SB equivalence}) are introduced by means of the Lagrange multiplier method (cf. Eqs. (4a) and (4b) of Ref. \onlinecite{KR}). Subsequently, a saddle-point approximation (being essentially a mean-field procedure) is formulated. The last step, yields an incorrect non-interacting ($U = 0$) limit. To tackle the situation, $\hat{z}_{i\sigma}$ operators, appearing in (\ref{SB Hubbard Hamiltonian}), are replaced in a formally equivalent form by making the multiplicative adjustment
\begin{equation}
\hat{z}_{i\sigma} \to \hat{\tilde{z}}_{i\sigma} \equiv (1 - d^{\dag}_{i} d_{i} - p^{\dag}_{i\sigma} p_{i\sigma})^{-1/2}\hat{z}_{i\sigma}(1 - e^{\dag}_{i} e_{i} - p^{\dag}_{i\bar{\sigma}} p_{i\bar{\sigma}})^{-1/2}.
\label{SB modified z factors}
\end{equation}
It is usually assumed also that all the Bose fields and Lagrange multipliers are site independent.
After all those steps are taken, the saddle-point grand potential functional $\mathcal{F} \equiv \mathcal{F}^{(SB)}$ is obtained in the form
\begin{eqnarray}
 \mathcal{F}^{(SB)} & = & -\frac{1}{\beta} \sum_{ \mathbf{k} \sigma } \ln (1 + e^{-\beta E^{(SB)}_{\mathbf{k} \sigma} }) + \Lambda U d^2 \nonumber \\ & + & \Lambda \lambda^{(1)} ( \sum_{ \sigma } p^{2}_{ \sigma} + e^{2} + d^{2} - 1) - \Lambda \sum_{ \sigma } \lambda^{(2)}_{ \sigma} ( p^{2}_{ \sigma} + d^{2}),
\label{SB SP MF GP}
\end{eqnarray}
with the effective quasiparticle energies
\begin{equation}
E^{(SB)}_{\mathbf{k} \sigma} = q_{\sigma} \epsilon_{\mathbf{k} \sigma} - \mu - \sigma h + \lambda^{(2)}_{ \sigma}, ~~~~~~~q_{\sigma} = \langle \hat{\tilde{z}}^{\dag}_{i\sigma} \hat{\tilde{z}}_{j\sigma} \rangle \equiv z_\sigma^2.
\label{SB E_k}
\end{equation}
In the expression for $ \mathcal{F}^{(SB)}$, the constraints (\ref{SB completeness}), (\ref{SB equivalence}) are understood in the average sense only, i.e.
\begin{equation}
e^{2} + \sum_{\sigma} p^{2}_{ \sigma} + d^{2} = 1,
\label{SB completeness MF}
\end{equation}
\begin{equation}
\langle f^{\dag}_{i\sigma} f_{i\sigma} \rangle \equiv n_{\sigma} = p^{2}_{ \sigma} + d^{2}.
\label{SB equivalence MF}
\end{equation}
Alternatively, those MF constraints may be obtained from the conditions
\begin{equation}
 \frac{\partial \mathcal{F}^{(SB)}}{\partial \lambda^{(1)}} = 0, ~~~~~~ \frac{\partial \mathcal{F}^{(SB)}}{\partial \lambda^{(2)}_{\sigma}} = 0.
\label{SB constraint through derivatives}
\end{equation}
Note that $ \mathcal{F}^{(SB)}$ given by (\ref{SB SP MF GP}) can also be obtained without invoking explicitly the functional integral formalism. Namely, one may simply replace boson operators in Hamiltonian (\ref{SB Hubbard Hamiltonian}) (with $\hat{z}_{i \sigma} \to \hat{\tilde{z}}_{i\sigma}$) augmented with the constraints (\ref{SB completeness}), (\ref{SB equivalence}), i.e.
\begin{equation}
\hat{H} + \sum_i \lambda^{(1)}_{i}\hat{Q}^{(1)}_{i} + \sum_i \lambda^{(2)}_{i\sigma}\hat{Q}^{(2)}_{i\sigma},
\label{SB Hamiltonian with the constraints}
\end{equation}
by the corresponding average values, without caring much about the precise meaning of such averaging procedure (the saddle point approximation is one of the possible ways to introduce it). This leads to the following effective (renormalized) MF Hamiltonian
\begin{eqnarray}
\hat{H}_{SB} &=& \sum_{ij\sigma}  t_{ij} f^{\dag}_{i\sigma}f_{j\sigma} z^{\ast}_{i\sigma} z _{j\sigma}   + \sum_{ i \sigma } \lambda^{(2)}_{i \sigma} ( f^{\dag}_{i\sigma} f_{i\sigma} -  |p_{i \sigma}|^2   -   |d_{i}|^2) \nonumber \\ &+& \sum_{ i \sigma } \lambda^{(1)}_{i} ( |p_{i \sigma}|^2   +  |e_{i}|^2 +   |d_{i}|^2 - 1) +  U\sum_{i} |d_{i}|^2,
\label{SB MF Hamiltonian}
\end{eqnarray}
in which only fermionic degrees of freedom are regarded as operators, whereas the bosonic variables are treated as classical (and usually also as spatially homogenous). This phenomenologically motivated procedure leads to the same results as the saddle-point approximation. Moreover, it allows to establish a closer connection with GA, as discussed next. Note that the classical correspondants of the local auxiliary Bose fields can be regarded as their Bose-condensed amplitudes and this SB feature may lead to spurious phase transitions in those condensed "ghost" fields, invariably regarded as MF order parameters.

\subsection{Equivalence of slave boson and SGA approaches \label{Equi}}

The variational SB procedure is carried out with respect to $\lambda^{(1)}$, $\lambda^{(2)}$, $e$, $p_{\sigma}$, and $d^2$ starting from the functional expression (\ref{SB SP MF GP}). However, the value of $\mathcal{F}^{(SB)}$ does not depend on $\lambda^{(1)}$ once the constraint (\ref{SB completeness MF}) is fulfilled.  Also, one may use (\ref{SB completeness MF}) to eliminate $e^2$ and (\ref{SB equivalence MF}) to eliminate $p^2_{\uparrow}$ and $p^2_{\downarrow}$ in favor of $n_\sigma$. Consequently, $\mathcal{F}^{(SB)}$ is brought to the simpler form
\begin{equation}
\mathcal{F}^{(SB)}= \Lambda U d^2 -\frac{1}{\beta } \sum_{ \mathbf{k} \sigma } \ln [1 + e^{-\beta E^{(SB)}_{\mathbf{k} \sigma} }] - \Lambda \sum_{ \sigma } \lambda^{(2)}_{ \sigma} n_{ \sigma},
\label{SB SP MF GP simpler}
\end{equation}
with $q_{\sigma}$ and $E_{\mathbf{k} \sigma}$ given by (\ref{SB E_k}).
Using the last expression for $\mathcal{F}^{(SB)}$, the variational procedure is carried out with respect to $\lambda_\sigma^{(2)}$, $n_{\sigma}$ and $d^2$, in addition to the chemical potential ($\mu$) determination (with $n=n_\ua + n_\da$ fixed).

To show the equivalence of SB and SGA, we make the correspondence
\eq
\lambda^{(2)}_{\sigma} \leftrightarrow - \lambda_{n} - \sigma \lambda_m.
\eqx
We may also make the corresponding change of variables: $(n_{\uparrow}, n_{\downarrow}) \leftrightarrow (n, m)$. Under these changes, the expressions (\ref{SB SP MF GP simpler Gutzwiller0}) and (\ref{SB SP MF GP simpler}) are identical. Explicitly,
\begin{eqnarray}
    \mathcal{F}^{(SB)} & = & \Lambda U d^2 -\frac{1}{\beta } \sum_{ \mathbf{k} \sigma } \ln [1 + e^{-\beta E^{(SB)}_{\mathbf{k} \sigma} }] - \Lambda \sum_{ \sigma } \lambda^{(2)}_{ \sigma} n_{ \sigma} = \nonumber \\
           & = & \Lambda U d^2 -\frac{1}{\beta } \sum_{ \mathbf{k} \sigma } \ln [1 + e^{-\beta E^{(SB)}_{\mathbf{k} \sigma} }] - \Lambda \sum_{ \sigma } (- \lambda_{n} - \sigma \lambda_m) n_{ \sigma} = \nonumber \\
           & = & -\frac{1}{\beta } \sum_{ \mathbf{k} \sigma } \ln [1 + e^{-\beta E^{(SB)}_{\mathbf{k} \sigma} }] + \Lambda (\lambda_{n} n + \lambda_m m + U d^2) \equiv \mathcal{F}^{(SGA)},
\end{eqnarray}
with the quasiparticle energies (given by Eqs. (\ref{E_k Gutzwiller}) and (\ref{SB E_k})), transforming accordingly
\begin{eqnarray}
    E^{(SB)}_{\mathbf{k} \sigma} & = & q_{\sigma} \epsilon_{\mathbf{k} \sigma} - \mu - \sigma h + \lambda^{(2)}_{ \sigma} = \nonumber \\
           & = & q_{\sigma} \epsilon_{\mathbf{k} \sigma} - \sigma (h + \lambda_m) - \mu - \lambda_{n} \equiv E^{(SGA)}_{\mathbf{k} \sigma},
\end{eqnarray}
where $ E^{(SGA)}_{\mathbf{k} \sigma} $ is the quasiparticle energy obtained from SGA (cf. Sec. \ref{GA2}).
Consequently, the equilibrium values of all MF variables, and hence all predictions of both models, are also identical, as they are determined from $\mathcal{F}$ through the same variational procedure.

In connection with the above reasoning one has to note, that the problem of equivalence of SB and GA was examined before by Gebhard for the single-band Hubbard model \cite{Gebhard older}, as well as the periodic Anderson model \cite{Gebhard}. In \cite{Gebhard} the equivalence with different SB schemes is obtained for particular limits (namely, with Coleman \cite{SB Anderson model} for large orbital degeneracy $N$, as well as with Kotliar-Ruckenstein \cite{KR} for high lattice dimensionality $D$).
Finally, let us note that the equal status of spin-rotationally invariant SB formalism for the multiband Hubbard model \cite{Lechermann} (cf. the next Section) and the corresponding Gutzwiller approximation has been discussed by B\"{u}neman and Gebhard \cite{Buenemann Gebhard} from a different perspective. Nonetheless, similarly to Refs. \onlinecite{Buenemann Weber} and \onlinecite{Buenemann Weber Gebhard}, the self-consistency constraints were not introduced explicitly there, in contrast to both Ref. \onlinecite{Buenemann Gebhard Thul} and the present approach.

\section{Generalization: Slave-Boson and SGA method equivalence in multiorbital case} \label{sec:appGen}
The proof of the equivalence of SB method of Sec. \ref{SB} and the SGA approach of Sec. \ref{GA2} can be straightforwardly generalized to other SB approaches. Let us consider first the SB formalism of Ref. \onlinecite{SB Multi}, characterized by an incomplete (density-density terms only) form of Coulomb interactions. Below we outline such generalization before turning to the explicit analysis of SGA in the single-band case. Assume we have $M_{O}$ orbitals ($2 M_{O}$ spin orbitals), with $\sigma$ and $\kappa$ being the spin and orbital indices, respectively. The SB formalism introduces $2 M_{O} + 1$ constraints: one reflecting the completeness relation for SB fields and $ 2 M_{O}$ constraints generalizing Eq. (\ref{SB equivalence MF}). On the MF level they read
\begin{equation}
e^{2}_{i} + \sum_{\sigma \kappa} p^{2}_{i \sigma \kappa} + \ldots = 1,
\label{SB completeness MF multi}
\end{equation}
\begin{equation}
\langle f^{\dag}_{i\sigma \kappa} f_{i\sigma \kappa} \rangle \equiv n_{i \sigma \kappa} = p^{2}_{i \sigma \kappa} + \ldots,
\label{SB equivalence MF multi}
\end{equation}
in which by $ (\ldots) $ we denote other terms corresponding to the double and higher ($\leq 2 M_0$) occupancies. One may use (\ref{SB completeness MF multi}) to eliminate $e_{i}^2$, and the remaining $ 2 M_{O}$ Eqs. (\ref{SB equivalence MF multi}) to eliminate $p_{i\sigma \kappa}$ in favor of physical $n_{i\sigma \kappa}$ and probabilities of double, triple, etc. occupancies. Then, within the SB method the band-narrowing factors can be obtained in the corresponding GA form. Next, SGA can be formulated with the help of a single-particle Hamiltonian, supplemented for each site with $2 M_{O}$ constraints of the form
\begin{equation}
\hat{Q}_{i\sigma \kappa} = -  \lambda_{i\sigma \kappa}(f^{\dag}_{i\sigma \kappa} f_{i\sigma \kappa} - n_{i\sigma \kappa}).
\label{SB constraints multi}
\end{equation}
This allows to treat $n_{i\sigma \kappa}$ as variational parameters within SGA. Moreover, the corresponding generalized grand potential functional $\mathcal{F}$ takes again identical form in both the SGA and the SB methods. In effect, both approaches would become equivalent in a general multiple-orbital case.

The SGA approach can also be generalized to the situations with either spin rotationally-invariant slave-boson formalism \cite{Li, Fresard2, JS1995}, and/or to the multiband Hubbard Hamiltonian which contains also terms off-diagonal in the spin-orbital index \cite{Buenemann Gebhard, Lechermann}. In such situation, we have to ascribe slave-boson fields (and the corresponding operator constraint) to e.g. each of the $(2 M_{O})^2$ operators $f^{\dag}_{i \sigma  \kappa}f_{i\sigma^{\prime} \kappa^{\prime}}$ or to their linear combinations (cf. Eqs. (10)-(12) of Ref. \onlinecite{Li}). Again, on the mean-field level, the averages of the SB constraints allow us to eliminate the slave boson amplitudes corresponding to empty and singly-occupied configurations. Those SB amplitudes are replaced by averages of operators $f^{\dag}_{i \sigma  \kappa}f_{i\sigma^{\prime} \kappa^{\prime}}$, i.e. the components of a local, single-particle density matrix \cite{Seibold}
\begin{equation}
\rho_{ii}^{(\sigma'\kappa^{\prime})(\sigma \kappa)} \equiv \langle f^{\dag}_{i\sigma \kappa} f_{i \sigma' \kappa^{\prime}} \rangle. \label{eq:local density matrix}
\end{equation}
Within the present method, we have to reintroduce the operator constraint for each component of the single-particle density matrix. Explicitly, we add to the MF Hamiltonian the terms of type
\begin{equation}
-\lambda_{ii}^{(\sigma'\kappa^{\prime})(\sigma \kappa)} (f^{\dag}_{i\sigma \kappa} f_{i \sigma' \kappa^{\prime}} - \rho_{ii}^{(\sigma'\kappa^{\prime})(\sigma \kappa)}).
\label{eq:constraint for general SB}
\end{equation}
Again, the amplitudes of relevant two-, three-, etc. electron configurations do not correspond to average values of any single-particle operator appearing in the MF Hamiltonian, and therefore do not require any constraints. In turn, the presence of the constraints allows to treat matrix elements $\rho_{ii}^{(\sigma \kappa) (\sigma'\kappa^{\prime})}$ as variational parameters.

As the last step, the corresponding generalized grand potential functional $\mathcal{F}$ is constructed and its form is again identical in both the SGA and the SB methods. The Lagrange multipliers of the present approach (i.e. $\lambda_{ii}^{(\sigma \kappa) (\sigma'\kappa^{\prime})}$) are in one-to-one correspondence with those originating from the SB formalism. In result, both approaches are fully equivalent. Also, the SB method provides us with a hint of how to construct the corresponding SGA Hamiltonian, but the specific ingredients of the SB formalism (i.e. amplitudes of condensed Bose fields) disappear eventually from the approach.

\section{Numerical comparison of GA vs. SGA: two dimensional square lattice \label{sec:numerics}}

In this Section we analyze the single-band Hubbard model (\ref{Hubbard Hamiltonian}), describing an \textit{almost localized Fermi liquid} (ALFL) for the case of two-dimensional square lattice within the tight-binding approximation, and in the limits of both finite $U$ (with $d \neq 0$) and $U \ra \infty$ (with $d \equiv 0$). We compare the results of the SGA method with those of GA solved in a way described in Sec. \ref{sec:formalism} and show, that they may differ essentially. This analysis is presented mainly for illustratory purposes.

In the following we take into account the first two hopping integrals $t$ and $t'$ (with fixed ratio $t'/t = 0.25$). We choose $t$ as the energy unit. The quasiparticle respective dispersion relations in the two analyzed approaches are
\eqn
E^{(SGA)}_{\bk \sigma} & = & q_\sigma \epsilon_\bk - \sigma (h + \lambda_m) - \tilde{\mu}, \label{eq:dispME} \\
E^{(GA)}_{\bk \sigma} & = & q_\sigma \epsilon_\bk - \sigma h - \mu, \label{eq:dispS}
\eqnx
with the bare dispersion relation
\eq
\epsilon_\bk  = - 2 t (\cos{k_x} + \cos{k_y}) + 4 t' \cos{k_x} \cos{k_y},
\eqx
and $q_\sigma \equiv q_\sigma(d, n, m)$ given by (\ref{q_sigma Gutzwiller}) (with the corresponding change of variables).

\subsection{GA versus SGA}

On the example of GA for the Hamiltonian (\ref{Hubbard Hamiltonian}) in a non-zero Zeeman field, we illustrate next the differences between the above two methods of solving the MF model (which in the present case means determining the values of $m$, $\lambda_m$, $\lambda_n$, $\mu$, and $d$). First, we use the SGA method described in Sec. \ref{GA2} (equivalent to the SB method of Sec. \ref{SB}) and treat both the magnetization $m$ and particle number $n$ as variational parameters, i.e. solve the complete set of equations (\ref{derivative of mathcalF A, lambda}). This will be labeled as the \textit{var} solution. The other possibility is to determine the value of $m$ in a self-consistent, non-variational manner (referred to as the \textit{s-c} solution). This last solution corresponds to the GA approach of Sec. \ref{sec:formalism} and can be achieved by solving Eqs. (\ref{der mathF ksi explicite standard})-(\ref{der mathF lambda m explicite2}).

An analogical comparison of the differences between the corresponding \textit{var} and \textit{s-c} solutions has been carried out earlier for the renormalized mean-field theory (RMFT) of the $t$-$J$ model \cite{Jedrak Spalek}. We show that in the present (ALFL) case the differences between the two formulations are even more pronounced. Explicitly, the grand potential functional within the SGA method is given by
\begin{equation}
\mathcal{F}^{(SGA)}   =  -\frac{1}{\beta  } \sum_{ \mathbf{k} \sigma }  \ln [1 + e^{-\beta E^{(SGA)}_{\mathbf{k} \sigma} }]   +  \Lambda( \lambda _{n} n   +   \lambda _{m} m  + U d^2).
\label{SB SP MF GP simpler Gutzwiller}
\end{equation}
The necessary minimization conditions ($\partial \mathcal{F} / \partial x_i = 0$, with $x_i = n, m, d, \lambda_n, \lambda_m$) lead to the following set of five equations
\begin{eqnarray}
    \lambda_{n}  & = &  - \frac{1}{\Lambda } \sum_{ \mathbf{k} \sigma } \frac{\partial  q_{\sigma}}{\partial n} f(E_{\mathbf{k} \sigma})  \epsilon_{\mathbf{k}}, \label{der mathF n explicite} \\
    \lambda_{m} & = & - \frac{1}{\Lambda } \sum_{ \mathbf{k} \sigma } \frac{\partial  q_{\sigma}}{\partial m} f(E_{\mathbf{k} \sigma})  \epsilon_{\mathbf{k}}, \label{der mathF m explicite} \\
    d & = & - \frac{1}{2 \Lambda U}  \sum_{ \mathbf{k} \sigma } \frac{\partial  q_{\sigma}}{\partial d} f(E_{\mathbf{k} \sigma})  \epsilon_{\mathbf{k}},
\label{der mathF d explicite} \\
    n & = & \frac{1}{\Lambda}\sum_{ \mathbf{k} \sigma }   f(E_{\mathbf{k} \sigma}),
\label{der mathF lambda n explicite} \\
    m  & = & \frac{1}{\Lambda}\sum_{ \mathbf{k} \sigma } \sigma  f(E_{\mathbf{k} \sigma}).
\label{der mathF lambda m explicite}
\end{eqnarray}
Within the \textit{s-c} approach, we solve only Eqs. (\ref{der mathF d explicite})-(\ref{der mathF lambda m explicite}) and put $\lambda_n = \lambda_m \equiv 0$; hence $d$ is the only variational parameter. In that particular case, we use $E^{(GA)}_{\bk \sigma}$ of Eq. (\ref{eq:dispS}) instead of $E^{(SGA)}_{\bk \sigma}$ of Eq. (\ref{eq:dispME}).

We solve the (\textit{var}, \textit{s-c}) equations for the case of a finite square lattice of the size $\Lambda = \Lambda_x \Lambda_y = 512 \times 512$, with $t^{\prime} = 0.25 \, t$ for low temperature $T = 1/\beta = 0.002$. We assume that the band filling is $n = 0.97$ and the Hubbard parameter is $U=8$.

\begin{figure}[h]
\begin{center}
\includegraphics[angle=270, width=16.5cm]{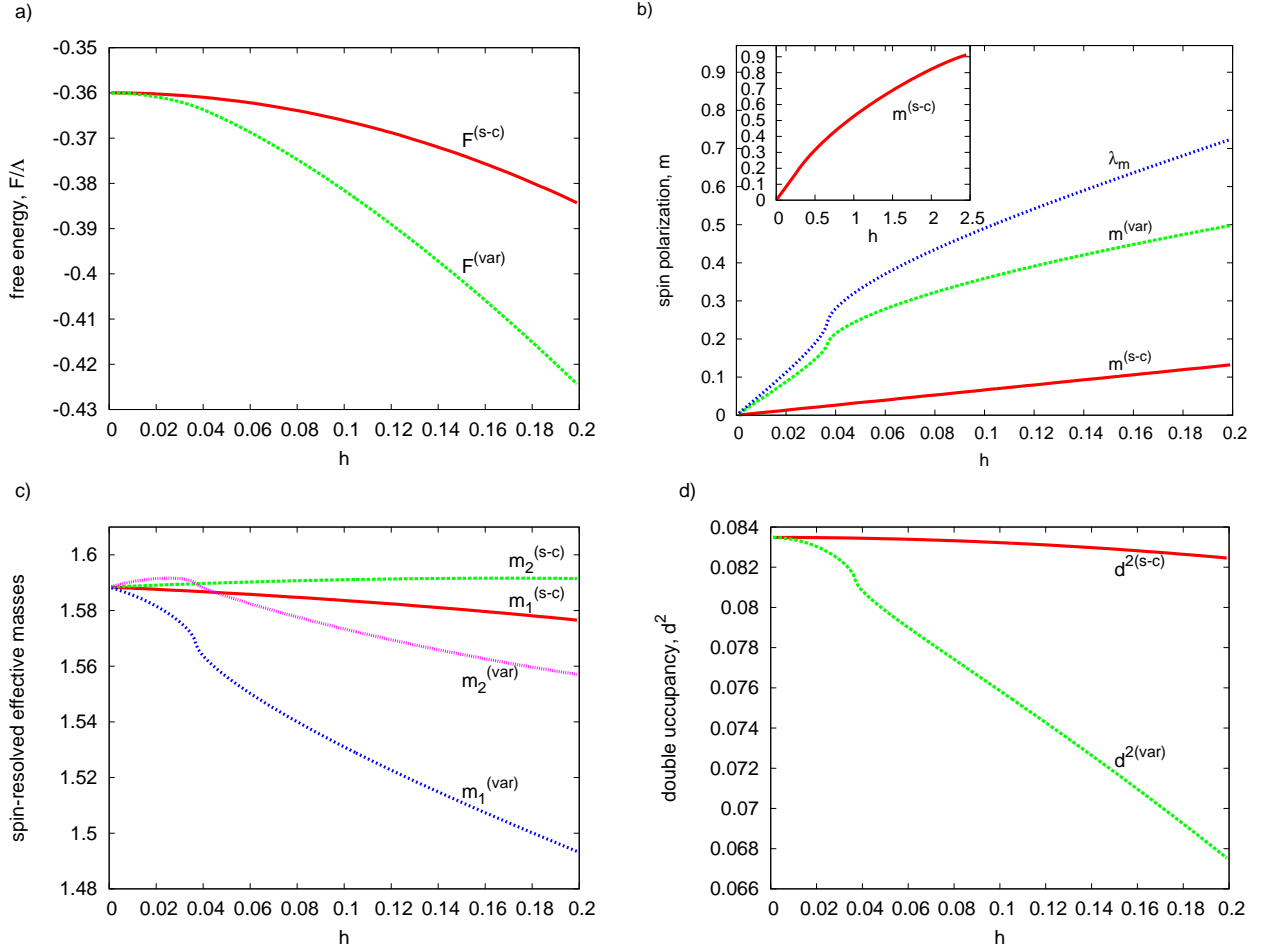}
\end{center}
\caption{(Color online) Magnetic field dependence of selected quantities for the band filling $n=0.97$. a) free energy; b) molecular field $\lambda_{m}$ and spin-polarization $m$; c) effective masses of quasiparticles $m_\sigma \equiv m_{1,2}$, and d) double occupancy probability for both \textit{var} and \textit{s-c} methods in each case. The free energy in the SGA method is smaller than that obtained by the GA of Sec. \ref{sec:formalism}. Note also a similar $h$-dependence of $m$ and $\lambda_{m}$, and large differences between $m^{(var)}$ and $m^{(s-c)}$ ($\Delta m^{(var)} \gtrsim 4 \Delta m^{(s-c)}$, with $\Delta m \equiv m_2 - m_1$). A metamagnetic behavior is observed around $h \approx 0.03$ only in the case of SGA method. Inset in b) shows an overall behavior of the spin-polarization in strong fields obtained within the \textit{s-c} scheme. }
\label{fig0}
\end{figure}

We can analyze the system behavior as a function of the Zeeman field $h$, as displayed in the panel composing Fig. \ref{fig0}a-d. From Fig. \ref{fig0}a we see that within the \textit{var} approach the free energy (which for such low $T$, is practically equal to the ground-state energy of the MF Hamiltonian (\ref{MF Gutzwiller Hamiltonian})) is essentially lower than that obtained within the \textit{s-c} treatment. This means that the probability distribution obtained within {\it s-c} scheme is not the optimal one from the point of view of MaxEnt inference. This fact does not necessarily mean that the {\it var} solution is closer to the exact ground-state energy than the {\it s-c} solution, as the Bogoliubov-Feynman inequality does not hold for GA.
In Fig. \ref{fig0}b we plot the field dependence of the molecular field $\lambda_{m}$ and magnetization $m$. The differences in the values of magnetization (hence also the susceptibility, not displayed here) calculated within both approaches, are large. Only for the \textit{var} method the nonlinear dependence of magnetization is observed, with a metamagnetic behavior appearing around $h \approx 0.03$. This behavior is caused by the strong molecular field $\lambda_m$, not present in the \textit{s-c} method and corresponds to the van Hove singularity of $\sigma = \da$ band moving across the Fermi level. Obviously, due to the presence of $\lambda_m$ the \textit{var} method is more sensitive to the details of the dispersion relation around the Fermi surface (cf. also Fig. \ref{figDOS2}). Please note, that values of magnetization closely resemble the nonlinear $h$-dependence of $\lambda_m$, which is a few times larger than the applied field $h$. In contrast, $m(h)$ is quite typical for the case of the \textit{s-c} approach (cf. also inset in Fig. \ref{fig0}b). The quasiparticle masses in the spin-subbands: $m_1 \equiv m_\ua = 1/q_\ua$ and $m_2 \equiv m_\da = 1/q_\da$, are exhibited in Fig. \ref{fig0}c. The mass splitting $ m_2 - m_1$ in the {\it var} method is a few times larger than for {\it s-c} because of the enhancement of the magnetization in the former case. Larger $m_2 - m_1$ is closer to the experiment \cite{McCollam}, which gives up to $m_2/m_1 \approx 4$. A decrease in both $m_1^{(var)}$ and $m_2^{(var)}$ observed for $h > 0.03$ is peculiar, as usually $m_2$ increases with the increasing field \cite{Korbel, JS2006}. However, in strong fields $m_2^{(var)}$ starts to increase around $h \approx 0.35$ and the high-field limit of large $m_2$ is properly recovered. For completeness, we present in Fig. \ref{fig0}d the applied field dependence of the double occupancy probability $d^2$. These results demonstrate that the present SGA method provides not only quantitative, but also qualitative differences as compared to GA method solved in the way introduced in Sec. \ref{sec:formalism}.
\begin{center}
Table I. Equilibrium values of chemical potentials, thermodynamic potentials (per site), mean-field variables and related quantities, for $h = 0.05$, both for \textit{var} and \textit{s-c} solutions.
$\tilde{\mu} $ stands for $\lambda_n + \mu $.\\
\begin{tabular}{|c|c c || c |c c|}\hline \hline
Variable & \textit{var} & \textit{s-c} & Variable & \textit{var} & \textit{s-c} \\ \hline
$\lambda_{n}$           & -3.6728 	& 0.0000    & $\lambda_{m}$  & 0.3318 		 & 0.0000 \\
$\mu$ 		            & 3.3685 	& -0.4062   & $ m $ 		 & 0.2516 	 & 0.0335 \\
$\tilde{\mu}$           & -0.3043 	& -0.4062   & $ d^2 $        & 0.0798 	 & 0.0834 \\
$ \Omega $              & -3.6334   & 0.0325   & $ n_{\uparrow} n_{\downarrow} $ & 0.2194  & 0.2349 \\
$\Omega - \lambda_n n$  & -0.0708   & 0.0325    & $ q _{\uparrow} $ & 0.6426 & 0.6304 \\
$ F $                   & -0.3660 	& -0.3615   & $ q _{\downarrow} $ & 0.6309 			 & 0.6289 \\ \hline \hline
\end{tabular}
\end{center}
Next, we analyze in detail the situation for fixed $h = 0.05$. The values of relevant quantities for this case are provided in Table I. Even for this relatively low value of the applied field, $d^2$ is slightly smaller within \textit{var} treatment, which indicates that the effect of strong electron correlations is, on the mean-field level, slightly enhanced in the SGA/SB approach, as compared to that in the GA (\textit{s-c}) treatment.

\subsection{Supplement: Strongly-correlated regime: the $U \ra \infty$ limit}

We discuss next the system properties in the limit of $U \ra \infty$ ($d=0$). In this case the Gutzwiller factors take the form
\eq
q_\sigma(n, m) = \frac{1-n}{1-n_\sigma}. \label{eq:Gutzd0}
\eqx

\begin{figure}
\begin{center}
\includegraphics[height=15cm]{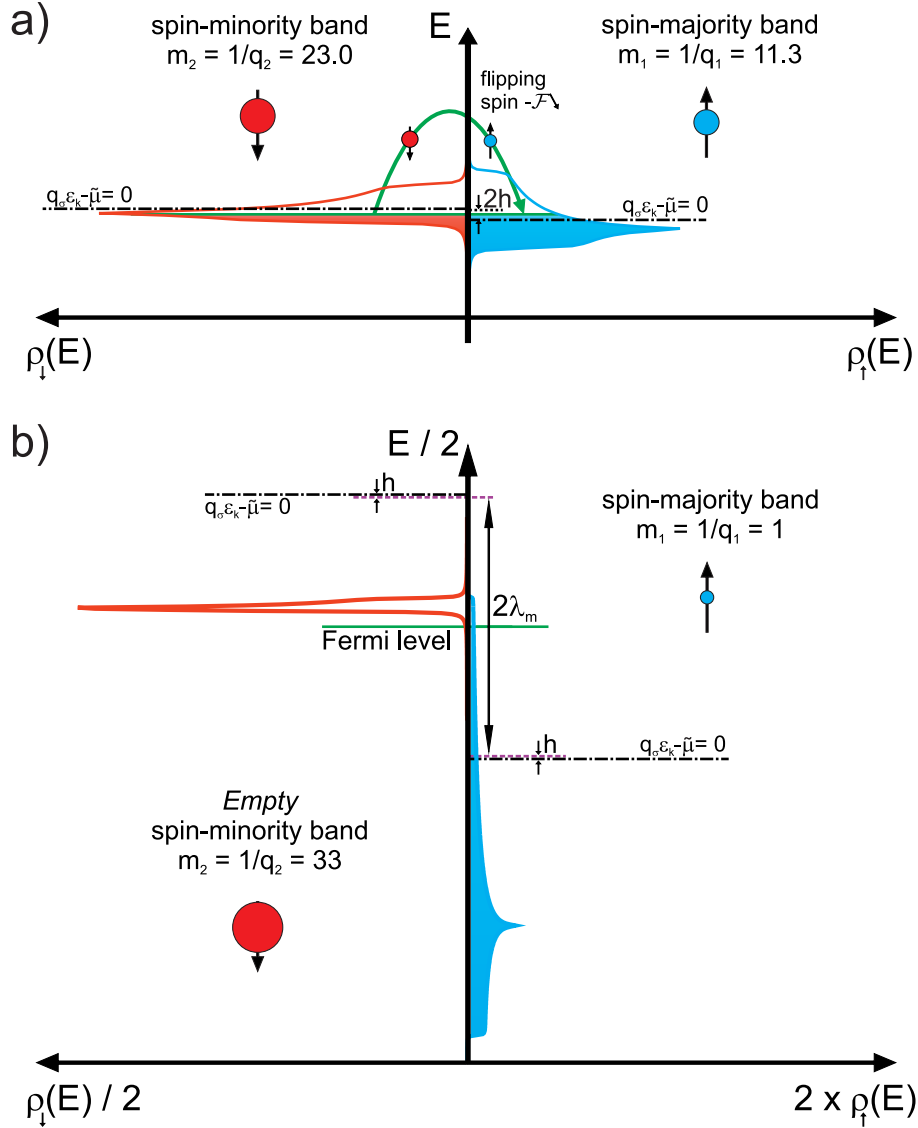}
\end{center}
\caption{(Color online) Density of states in the limit of $U \ra \infty$ ($d=0$) for the spin-majority ($\sigma = \ua$) and the spin-minority ($\sigma = \da$) subbands, for both GA: (a) and SGA (b). The dot-dashed lines show the reference energy (defined by $q_\sigma \epsilon_\bk - \tilde{\mu} = 0$). Note that the axes in b) are scaled for clarity. The calculations were performed for $h=0.05$. The ground state in b) is that of saturated ferromagnet ($m = n$).}
\label{figDOSd0}
\end{figure}

It turns out that the saturated ferromagnetic solution ($m=n$) is the ground state for the SGA method, whereas GA approach provides a ferromagnetic ground state ($0<m<n$). The density of states for both methods is exhibited in Fig. \ref{figDOSd0}.

To understand the reason behind the ferromagnetic ground state in SGA (or equivalent SB) method it is useful to recall the physical meaning of $\lambda_m$. Namely, this parameter optimizes the free energy by allowing for a mismatch between chemical potentials of the spin-subbands (cf. Fig. \ref{figDOS2}b). It turns out that in the limit of $d=0$ it is beneficial for one subband to be completely empty, while all electrons occupy the other one. This is easy to understand as in such situation one of the bands becomes very broad (acquires the bare bandwidth value as $q_\ua = 1$). Such broad band is favorable, as then its "center-of-gravity" shifts to negative energies. This ferromagnetic behavior is present under any non-zero Zeeman field $h$.
Parenthetically, in the $t$-$J$ and $t$-$J$-$U$ models ferromagnetism is strongly suppressed by the $J \sum_{<ij>} \mathbf{S_i} \cdot \mathbf{S_j}$ term, which favors antiferromagnetism. On the other hand, the presence of the saturated ferromagnetism for $n \rightarrow 1$ is in agreement with the Nagaoka theorem \cite{Nagaoka} (cf. also \cite{Federico Becca and Sandro Sorella, Hyowon Park}).

\begin{figure}
\begin{center}
\includegraphics[angle=270, width=16.5cm]{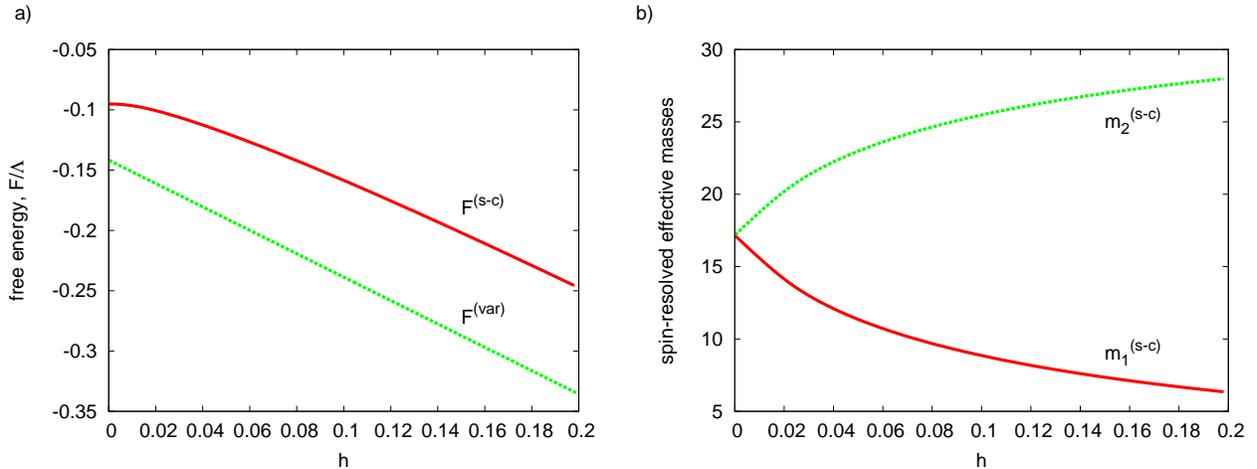}
\end{center}
\caption{(Color online) Magnetic field dependence of several quantities in the limit of $U \ra \infty$ and for typical $n$: a) free energy and b) effective masses $m_\sigma \equiv m_{1,2}$ of quasiparticles. The free energy in the SGA method is, as in Fig. \ref{fig0}, much smaller than that obtained by the standard GA. Note that $m_1^{(var)} = 1$ and $m_2^{(var)} = 1/(1-n)$. Note that the slope of $F^{(var)}$ is constant and nonzero even for $h \to 0^+$, which indicates that the ground state is saturated ferromagnetic even at $h=0$, since $\partial F^{(var)} / \partial h = - m = -n$. This is in contrast with the {\it s-c} method because then the corresponding relation is not fulfilled, $\partial F^{(s-c)} /~\partial~h~\neq~-~m$.}
\label{fig5}
\end{figure}

For completeness we present in Fig. \ref{fig5} the field dependence of the free energy and quasiparticle masses obtained within both GA and SGA approaches. Masses are exhibited only for the the GA approach, as in the SGA scheme they are not dependent on the Zeeman field (the ground state is that of saturated ferromagnet) and equal to $m_1^{(var)} = 1$, $m_2^{(var)} = 1/(1-n) \approx 33$. Again, the free energy obtained in the \textit{var} scheme is lower than the corresponding energy in the \textit{s-c} approach (here even at $h \ra 0$, as the ferromagnetic instability is present for arbitrarily low field $h$).

A brief relation of our concept of ALFL to the original Landau theory \cite{Landau} can be made. First, we have in both approaches the mass renormalization factor. Second, we have here the two effective correlation-induced fields $\lambda_m$ and $\lambda_n$, one responsible for magnetism enhancement ($\lambda_m$) and the other ($\lambda_n$) for the chemical potential shift. The effective fields and the mass enhancement are calculated here explicitly as a function of the microscopic parameter $U/t$ and the band filling $n$ within the variational procedure, whereas in the Landau theory they are expressed in terms of the phenomenological parametrization through $F_l^s$ and $F_l^a$ \cite{Abrikosov} representing the interparticle interaction for electrons near the Fermi surface.

\section{Critical overview and further extensions} \label{sec:extend}

The SGA method based on combining GA with the MaxEnt principle is applicable to every Gutzwiller scheme, not necessarily restricted to the Hubbard model. Namely, the method can be applied, among others, to study periodic Anderson \cite{Rice}, $t$-$J$ \cite{Marcin Letters, Marcin 1, Jedrak Spalek, Edegger}, $t$-$J$-$U$ \cite{Yuan, Heiselberg}, and multiband Hubbard \cite{Buenemann Weber, Buenemann Weber Gebhard, BunemannX} models used in the context of correlated electron systems, as well as for fermions in optical lattices \cite{Yuan, Heiselberg}. It can also be extended to the spin-rotationally invariant situation (cf. also Sec. \ref{sec:appGen}). Below we outline how this could be achieved in selected cases.

\subsection{$t$-$J$-$U$ model for optical lattices}

In Ref. \onlinecite{Yuan} the authors apply Gutzwiller method to study the $t$-$J$-$U$ Hamiltonian. The Gutzwiller factors given by Eqs. (9) and (10) of this paper depend on the antiferromagnetic (staggered) magnetization denoted as $m$ and defined by the self-consistent equation (16). Next, the ground-state energy (11) is minimized with respect to $m$ and other parameters, which yields Eq. (21) being in contradiction with the self-consistent equation (16). Eq. (21) is solved next along with Eqs. (19), (20), (22), and (23). In view of our results, such procedure should be modified, because Eqs. (16) and (21) are contradictory (the magnetization obtained from Eqs. (19)-(23) is different from that calculated for the same parameters from Eq. (16)). To fix this the Hamiltonian (13) should be supplemented with the appropriate constraints
\eq
\hat{H}_{MF} \ra \hat{H}_\lambda = \hat{H}_{MF} - \sum_i \lambda_{m, i} (\hat{m}_i - m_i) - \lambda_n \sum_i (\hat{n}_i - n), \label{eq:Hchange}
\eqx
where $\hat{m}_i \equiv (c_{i\ua} \dg c_{i\ua} - c_{i\da} \dg c_{i\da})$, $m_i = \langle \hat{m}_i \rangle$, and $m \equiv (-1)^i m_i$. This would assure consistency of Eq. (16) with the appropriate minimization condition $ \langle \partial \hat{H}_\lambda/ \partial m \rangle = 0$. Also, the equations $ \langle \partial\hat{H}_\lambda / \partial \lambda_m \rangle = 0$ (with $\lambda_m = \lambda_{m, i} (-1)^i$) and $\langle \partial \hat{H}_\lambda/ \partial \lambda_n \rangle = 0$ should have to be solved concomitantly. Such procedure would yield altered results than those presented in that paper.

In Ref. \onlinecite{Heiselberg} the author analyzes the $t$-$J$-$U$ model for fermions in optical lattice. Gutzwiller factors given by Eqs. (2) and (3) depend again on the staggered magnetization $m$ defined via the self-consistent equation (7). Next, variational energy (8) is invoked and the corresponding free energy is minimized with respect to some parameters, yielding Eqs. (9)-(11). However, for obtaining equation for $m$, the author refers to the paper on slave-bosons \cite{YuanSB}. Using the SGA scheme the Hamiltonian (1) would be supplemented by the constraints as in our Eq. (\ref{eq:Hchange}). This would allow for including the condition $\partial \mathcal{F}/\partial m = 0$, in accordance with the self-consistent equation (7) and without the necessity of invoking the slave-boson formalism. Moreover, the approach of \cite{Heiselberg} does not include the Lagrange multipliers coming from the SB method. Therefore, it is not clear how the method of \cite{Heiselberg} relates to SB.

\subsection{Gutzwiller approximation in the multiband case} \label{sec:relation:multiGutz}
In Refs. \onlinecite{Buenemann Weber} and \onlinecite{Buenemann Weber Gebhard} the authors study, among others, magnetic properties and introduce the spin-dependent Fermi level $E_{F\sigma}$ (cf. Eqs. (44) and (49) of Ref. \onlinecite{Buenemann Weber Gebhard}). Consequently, in those papers, as well as in \cite{Vollhardt} for the single-band Hubbard model, the scalar term ($\sim m \lambda_m$) involving magnetization $m$ is absent when compared to our approach. Also, magnetization $m$ (or equivalently $n_\sigma$) is fixed at a constant value, i.e. not treated as a variational parameter. Therefore, for this case, our method (as well as the SB approach \cite{KR}) may lead to different results.
Interestingly, in \cite{Buenemann Gebhard Thul} the same authors introduce constraint terms for $n_\sigma$ (equivalent to those of the present method). This step is taken in order to construct the effective Hamiltonian and analyze its full spectrum (ground state and excited states). However, the constraints are also important for the ground-state analysis.

\subsection{Explicitly spin-rotational-invariant formulation of SGA method: general features} \label{sec:extendC}

In Refs. \onlinecite{Seibold, DiCiolo, Markiewicz} the authors use the spin-rotational-invariant Gutzwiller formalism with the Gutzwiller factors matrix $\mathbf{z_i}$ based on the spin-rotationally invariant SB approach \cite{Li, Fresard2, JS1995}. This formalism is founded on the generalized time-dependent Hartree-Fock theory \cite{Nuclear many-body problem}, and we believe that it is correct from the statistical-physics point of view. Alternatively, within the framework of our method appropriate extra Lagrange multipliers $\lambda_{ij}^{\sigma\sigma'}$, leading to the local terms of the type
\eq
- \sum_{\sigma \sigma^{\prime}}\lambda_{ii}^{\sigma\sigma'} (c_{i \sigma'} \dg c_{i\sigma} - \rho_{ii}^{\sigma\sigma'}), \label{eq:GRPA1}
\eqx
can be added to the MF Hamiltonian to allow for a variational treatment with respect to those variables. This addition is needed because the matrix  $\mathbf{z_i}$ depends on the mean-fields $\rho_{ij}^{\sigma\sigma'} \equiv \langle \phi|c_{j \sigma'} \dg c_{i\sigma} | \phi \rangle$.

Explicitly, in the present situation, the constraints (\ref{eq:GRPA1}) can be decomposed into the two following parts
\eq
 [\vec{\lambda}_{m, i} \cdot (\mathbf{\hat{S}}_i - \frac{1}{2}\vec{m} \cdot \mathbf{1}) - \lambda_{n,i} (\hat{n}_i - n)], \label{eq:SCrot}
\eqx
where $\vec{\lambda}_{m, i} \equiv (\lambda^x_{m, i}, \lambda^y_{m, i}, \lambda^z_{m, i})$ and $\lambda_{n,i}$ are the corresponding molecular fields, $\vec{m}~\equiv~(m_x, m_y, m_z)$ is the magnetization vector, and $\mathbf{1}$ is the $2 \times 2$ unity operator. The site spin operator is defined as
\eq
\mathbf{\hat{S}}_i = \frac{1}{2} \sum_{\sigma \sigma^{\prime}}{c_{i\sigma}^\dagger(\vec{\tau})_{\sigma \sigma'} c_{i\sigma'}},
\eqx
and $\vec{\tau} \equiv (\tau_x, \tau_y, \tau_z)$ is the vector of Pauli matrices. The decomposition (\ref{eq:SCrot}) follows from the fact that any $2 \times 2$ matrix $\hat{\lambda}_i$ in the spin $(1/2)$ space can be decomposed into the vector and the scalar parts, i.e.
\eq
\hat{\lambda}_i = \vec{\lambda}_{m,i} \cdot \vec{\tau} + \lambda_{n,i} \mathbf{1}.
\eqx
The four molecular fields should be determined variationally (cf. corresponding discussion \cite{JS1995}). In result, the effective Hamiltonian $\hat{H}_{\lambda}$ replacing the original one has also the spin-rotational-invariant form. Such model can be applied to the situations with noncollinear spin ordering, as well as when discussing the spatial fluctuations of the magnetic molecular field $\vec{\lambda}_{m, i}$.

\section{Conclusions} \label{sec:summary}

\subsection{Summary}

In the present paper we have proposed a new approach (SGA) to the Gutzwiller approximation (GA) for the Hubbard and related models. To solve the effective single-particle mean-field (MF) Hamiltonian in an optimal way, we employ the maximum-entropy (MaxEnt) method. We also prove that our treatment of GA is fully equivalent to the saddle-point slave boson (SB) approach. Additionally, the MaxEnt-based treatment of GA introduces extra terms (constraints), which we include within the Lagrange-multiplier method. The motivation for introducing such terms is different in those two approaches, but they have the same form in both. Consequently, SB MF method obeys (in a sense, accidentally) the requirements of the maximum-entropy inference applied to MF models. Thus, the presence of molecular fields providing an advantage of the SB MF treatment is shared, as we have shown, with GA, but only if the latter is combined with the proper MaxEnt treatment. Moreover, such additional terms are not only specific ingredient of the MF SB formalism, but rather a generic feature of the statistically-consistent treatment of the MF models. We have also illustrated why the basic method of solving GA fails (Sec. \ref{GA2}), as well as the new features of the SGA method on the example of the single-band Hubbard model.


\subsection{Outlook}

We should emphasize, our method offers also a possibility of going beyond the SB techniques. Namely, the presented SGA approach can easily be extended by incorporating more advanced schemes of calculating averages beyond the standard GA. Namely, such improved schemes (for the $U \to \infty$ limit) have been proposed recently by Fukushima \cite{Fukushima} (and applied by us \cite{Jedrak Spalek} to the $t$-$J$ model within the MaxEnt-based approach); alternative approaches were formulated in \cite{Sigrist, Ogata Himeda}. By improving the averaging procedure beyond GA one may obtain a solution of comparable quality to those of the VMC calculations (see \cite{Fukushima}, Figs. 3 and 4). Another generalization of the Gutzwiller scheme is the inclusion of $1/D$ corrections \cite{Buenemann Gebhard Thul}, where $D$ is the lattice dimensionality. We should be able to see progress along these lines in the near future.

\begin{acknowledgments}
The work was supported by Ministry of Higher Education and Science, Grants Nos. N N202 128736 and N N202 173735. The discussions with Prof. Krzysztof Byczuk and Prof. Maciej Maśka are appreciated. Technical help from Marcin Abram is also appreciated.
\end{acknowledgments}

\appendix

\section{Stoner (Hartree-Fock) limit} \label{appHF}
In this Appendix we discuss the (mean-field) Stoner model of uniformly magnetically polarized band electrons \cite{FB}, as it represents a generic example of the Hartree-Fock approach. We show that the MaxEnt method does not introduce any new features into a standard formulation, as should be the case. The Hubbard Hamiltonian in the reciprocal space is expressed as
\begin{equation}
\hat{H} = \sum_{\mathbf{k} \sigma} \xi_{\mathbf{k}\sigma} c^{\dag}_{\mathbf{k} \sigma} c_{\mathbf{k} \sigma} + \frac{U}{2 \Lambda} \sum_{\mathbf{k} \mathbf{k}^{\prime} \mathbf{q}, \sigma}
 c^{\dag}_{\mathbf{k}+ \mathbf{q}, \sigma} c^{\dag}_{\mathbf{k}^{\prime} - \mathbf{q}, \sigma^{\prime}}c_{\mathbf{k}^{\prime} \sigma^{\prime}}c_{\mathbf{k} \sigma}
\label{Parent Hubbard}
\end{equation}
Here $\Lambda$ is the volume of the system (number of sites),  $\xi_{\mathbf{k}\sigma} = \epsilon_{\mathbf{k}} - \sigma h - \mu $, and $\epsilon_{\mathbf{k}}$ is the single- particle bare dispersion relation. $U$ is a matrix element of the Coulomb interaction, assumed to be contact in the real space. Applying the Hartree-Fock approximation we assume that
\begin{equation}
\langle c^{\dag}_{\mathbf{k}^{\prime} \sigma^{\prime}}c_{\mathbf{k} \sigma} \rangle  = \delta_{\mathbf{k}\mathbf{k}^{\prime}} \delta_{\sigma \sigma^{\prime}} n_{\mathbf{k}\sigma},
\end{equation}
and eventually we put  $n_{\sigma} = \frac{1}{\Lambda} \sum_{\bk} n_{\bk \sigma}$.
Consequently, the starting MF Hamiltonian reads
\begin{eqnarray}
\hat{H}_{HF} &=& \sum_{\mathbf{k} \sigma} (\xi_{\mathbf{k}\sigma} + U n_{\bar{\sigma}}) c^{\dag}_{\mathbf{k} \sigma} c_{\mathbf{k} \sigma} - \frac{U \Lambda}{2} \sum_{\sigma \sigma^{\prime}} n_{\sigma} n_{\sigma^{\prime}} +  \frac{U \Lambda}{2} \sum_{\sigma } n^{2}_{\sigma} \nonumber \\
    &= & \sum_{\mathbf{k} \sigma} \big(\epsilon_{\mathbf{k}} + \frac{U n}{2} - \sigma (h + \frac{U m}{2}) - \mu\big)c^{\dag}_{\mathbf{k} \sigma} c_{\mathbf{k} \sigma} - \frac{U\Lambda}{4}\big(n^2 - m^2),
\label{MF Hamiltonian Stoner}
\end{eqnarray}
with $n = \sum_{\sigma} n_{\sigma}$, $m = \sum_{\sigma} \sigma n_{\sigma}$. The next step is the addition of global constraint terms for $n$ and $m$,
\begin{equation}
\hat{Q}_{\lambda} = -\lambda_{n} \big ( \sum_{\mathbf{k} \sigma}  c^{\dag}_{\mathbf{k} \sigma} c_{\mathbf{k} \sigma} -  \Lambda n\big)  -\lambda_{m} \big ( \sum_{\mathbf{k} \sigma}  \sigma c^{\dag}_{\mathbf{k} \sigma} c_{\mathbf{k} \sigma} -  \Lambda m \big).
\label{constraints on MF Hamiltonian Stoner}
\end{equation}
This step yields
\begin{equation}
\hat{H}_{\lambda} \equiv \hat{H}_{HF} +  \hat{Q}_{\lambda} = \sum_{\mathbf{k} \sigma} E_{\mathbf{k}\sigma} c^{\dag}_{\mathbf{k} \sigma} c_{\mathbf{k} \sigma} + \Lambda \big[ \big(\lambda_n n + \lambda_m m\big) - \frac{U}{4}\big(n^2 - m^2) \big].
\label{MF Hamiltonian Stoner with constraints}
\end{equation}
In the above formula $ E_{\mathbf{k}\sigma} = \epsilon_{\mathbf{k}} -\lambda_n+ \frac{1}{2}Un - \sigma (h + \lambda_m + \frac{1}{2}U m) - \mu$. Using (\ref{MF Hamiltonian Stoner with constraints}) the (generalized) grand potential functional $\mathcal{F}(n, m, \lambda_n, \lambda_m)$ can be easily found
\begin{equation}
\mathcal{F}^{(HF)} =  -\beta^{-1} \sum_{\mathbf{k} \sigma} \ln (1 + e^{-\beta E_{\mathbf{k}\sigma}} ) + \Lambda \big[\big(\lambda_n n + \lambda_m m\big) - \frac{U }{4}\big(n^2 - m^2) \big].
\label{F functional for Hamiltonian Stoner}
\end{equation}
For the equilibrium situation the following equations
\begin{equation}
\frac{\partial \mathcal{F}^{(HF)}}{\partial n} = \Lambda(\lambda_n - \frac{1}{2} U n) + \frac{1}{2} U\sum_{\mathbf{k} \sigma} f (E_{\mathbf{k}\sigma})  = 0,
\label{der w.r.t. n of F functional }
\end{equation}
\begin{equation}
\frac{\partial \mathcal{F}^{(HF)}}{\partial m} = \Lambda(\lambda_m + \frac{1}{2} U m) -\frac{1}{2} U\sum_{\mathbf{k} \sigma} \sigma f (E_{\mathbf{k}\sigma})  = 0,
\label{der w.r.t. m of F functional }
\end{equation}
\begin{equation}
\frac{\partial \mathcal{F}^{(HF)}}{\partial \lambda_n} = \Lambda n - \sum_{\mathbf{k} \sigma}f(E_{\mathbf{k}\sigma})  = 0,
\label{der w.r.t. lambda1 of F functional }
\end{equation}
\begin{equation}
\frac{\partial \mathcal{F}^{(HF)}}{\partial \lambda_m} = \Lambda m - \sum_{\mathbf{k} \sigma} \sigma f(E_{\mathbf{k}\sigma})  = 0,
\label{der w.r.t. lambda2 of F functional }
\end{equation}
 must be simultaneously obeyed. It is easy to see that then $\lambda_n = \lambda_m = 0$, in accordance with the general  discussion on  Hartree-type of MF Hamiltonians \cite{JJJS_arx_0}, and we are left with the standard self-consistent equations (\ref{der w.r.t. lambda1 of F functional }) and (\ref{der w.r.t. lambda2 of F functional }) with $E_{\bk \sigma} = \epsilon_\bk + \frac{1}{2} U (n - \sigma m) - \sigma h - \mu$. One should underline again that the standard situation arises because $q_\sigma \equiv 1$.

{}

\end{document}